# Development and validation of the Converging Lenses Concept Inventory for middle school physics education

Salome Wörner[1,2]*, Sebastian Becker[3], Stefan Küchemann[2], Katharina Scheiter[1,4], and Jochen Kuhn[5]

[1] Leibniz-Institut für Wissensmedien, 72076 Tübingen, Germany

[2] Technische Universität Kaiserslautern, 67663 Kaiserslautern, Germany

[3] University of Cologne, 50923 Köln, Germany

[4] University of Potsdam, 14476 Potsdam, Germany

[5] LMU Munich, 80333 München, Germany

*Correspondence concerning this article should be addressed to Salome Wörner, Leibniz-Institut für Wissensmedien, Schleichstraße 6, D-72076 Tübingen. Email: s.woerner@iwm-tuebingen.de

Author Note

Salome Wörner, Leibniz-Institut für Wissensmedien Tübingen, Schleichstr. 6, 72076 Tübingen, Germany; s.woerner@iwm-tuebingen.de, ORCID: 0000-0002-7049-9316

Sebastian Becker, Digital Education Research Group, Faculty of Mathematics and Natural Sciences, University of Cologne, Herbert-Lewin-Str. 10, 50931 Köln, Germany; sbeckerg@uni-koeln.de, ORCID: 0000-0002-2461-0992

Stefan Küchemann, Dpt. Physics/Physics Education Research Group, Technische Universität Kaiserslautern, Erwin-Schrödinger-Straße 46, 67663 Kaiserslautern, Germany; s.kuechemann@physik.uni-kl.de, ORCID: 0000-0003-2729-1592

Katharina Scheiter, Leibniz-Institut für Wissensmedien Tübingen, Schleichstr. 6, 72076 Tübingen, Germany; University of Potsdam, Educational Science Department, Karl-Liebknecht-Str. 24-25, 14476 Potsdam; katharina.scheiter@uni-potsdam.de, ORCID: 0000-0002-9397-7544

Jochen Kuhn, Chair of Physics Education/Faculty of Physics, LMU Munich, Theresienstr. 37, 80333 München; jochen.kuhn@lmu.de, ORCID: 0000-0002-6985-3218

**Abstract**

Optics is a core field in the curricula of secondary physics education. In this study, we present the development and validation of a test instrument in the field of optics, the Converging Lenses Concept Inventory (CLCI). It can be used as a formative or a summative assessment of middle school students' conceptual understanding of image formation by converging lenses. The CLCI assesses: (1) the overall understanding of fundamental concepts related to converging lenses, (2) the understanding of specific concepts, and (3) students' propensity for difficulties within this topic. The initial CLCI consists of 16 multiple-choice items; however, one item was removed based on various quality checks. We validated the CLCI thoroughly with distractor analyses, classical test theory, item response theory, structural analyses, and analyses of students' total scores at different measurement points as quantitative approaches, as well as student interviews and an expert survey as qualitative approaches. The quantitative analyses are mostly based on a dataset of $N = 318$ middle school students who took the CLCI as a posttest. The student interviews were conducted with seven middle school students after they were taught the concepts of converging lenses. The expert survey included five experts who evaluated both individual items and the test as a whole. The analyses showed good to excellent results for the test instrument, corroborating the 15-item CLCI's validity and its compliance with the three foci outlined above.

## I. INTRODUCTION

Optics is one of the core fields in physics [1–3] and, as such, it is universally represented in the curricula of secondary physics education across the world (e.g., in K-12 education [4]). One fundamental topic in optics education is the learning of concepts regarding the refraction of light by a converging lens and how images can be formed using a lens. This phenomenon of an illuminated object being refracted by a lens and projected on a screen is depicted in Fig. 1.

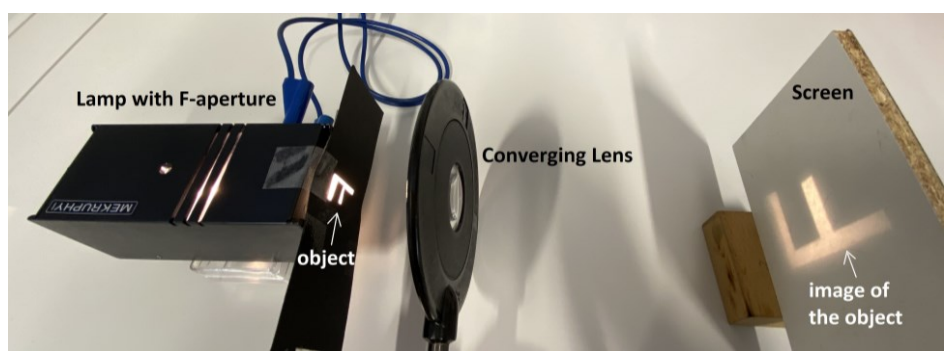

FIG. 1. An illuminated object being refracted by a lens and projected on a screen.

Conceptual understanding is a fundamental aim of science education [4–6]. It can be described as relational knowledge about the key concepts of a domain and the interrelations among these concepts, including the understanding of relations between observable phenomena and the underlying abstract and invisible principles [7,8].

In this article, we introduce a test instrument that can be used to assess middle school students' conceptual understanding of image formation by a converging lens. This instrument is called the Converging Lenses Concept Inventory, or CLCI. A concept inventory (CI) contains conceptual ideas and common students' difficulties in the single items and is easily usable due to its multiple-choice item format. In this paper, we describe the development and validation process of the instrument. We validated the CLCI thoroughly with multiple complementary approaches, namely, quantitative analyses including distractor analyses, classical test theory, item response theory, structural analyses, and analyses of students' total scores at different measurement points, as well as qualitative analyses including student interviews and an expert survey.

In the assessment of the CLCI's validity, we followed the recommendations of Jorion and colleagues [9]. They suggested an analytical framework to validate three claims that developers frequently make about the potential of a CI to successfully assess students' conceptual understanding. These "three generic claims underlie the common uses of CIs in classrooms and educational research:

(1) Students' CI scores can be used to indicate their overall understanding of all concepts identified in the CI. […]

(2) Students' CI scores can be used to indicate their understanding of specific concepts. […]

(3) Students' CI scores can be used to indicate their propensity for misconceptions or student errors. […]" [9] (p. 456)

We aim to make evidence-based statements about the validity of the CLCI concerning the degree to which this test instrument complies with these three claims.

## II. BACKGROUND

### A. Geometrical optics and lenses in middle school

The diffraction of light at the interface between optical components, such as lenses and the air, can be described by Snell's law. Snell's law relates the refractive indexes of the two materials on each side of the interface to the refractive angle [1,2]. This implies that light is refracted, that is, it changes its direction at each interface where the light enters a medium with a different refraction index. Snell's law is often excluded from the curricula of high and middle schools. Instead, students learn a method for drawing ray diagrams for objects placed at various locations in front of, for example, a double convex lens (converging lens). To draw a ray diagram (see Fig. 2), the three rules of refraction for a double convex lens are applied as follows:

- Any incident ray traveling parallel to the principal axis of a converging lens will refract through the lens and travel through the focal point on the opposite side of the lens (parallel ray).

- Any incident ray that passes through the center of the lens will in effect continue in the same direction that it had when it entered the lens (central ray).

- Any incident ray traveling through the focal point on the way to the lens will refract through the lens and travel parallel to the principal axis (focal-point ray).

We call the parallel ray, central ray, and focal-point ray the three construction rays.

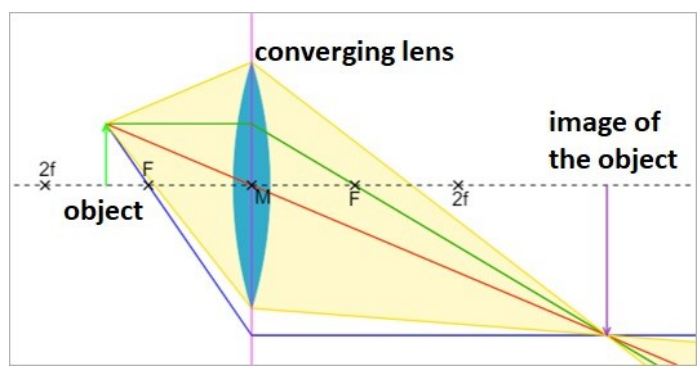

FIG. 2. Conceptual idea of the refraction and imaging process drawn in a ray diagram according to the three rules of refraction. The three construction rays predict where a point of the object is projected (parallel ray in green, central ray in red, and focal-point ray in blue).

Figure 2 shows a schematic view of an illuminated object (in the left), which is refracted by a converging lens (middle), such that on the right, the image of the object is formed. When students in middle or secondary school are taught about optics in their physics classes, typically, several phenomena of image formation by a converging lens are taught. These are the influence of the object distance, object height, and focal length of the lens on the resulting image of the object. Moreover, the effects of partly covering the lens and removing the screen are additional topics that are addressed in class.

## B. Students' difficulties with image formation by converging lenses

Due to the ubiquitous nature of light and refraction phenomena, students hold several conceptions that are inconsistent with the actual phenomena and their physical descriptions (often termed "misconceptions," cf. [10], "students' conceptions" or "alternative conceptions," cf. [11], or "naïve conceptions," cf. [12]). Together with the procedure of the construction of an optical image using the construction rays, this may lead to difficulties for students during learning. Galili and colleagues [13] (p. 280) introduce five core concepts (CC) about image formation and observation in the context of converging lenses:

> (CC1) "Object point and image point: An extended object is an assembly of object points; an image is an assembly of image points."
>
> (CC2) "Optical device: The entire flux of light diverging from each object point and interacting with the optical component contributes to the image point."
>
> (CC3) "Diverging, converging light fluxes: An optical device may change flux of light diverging from an object point into another flux that either (a) converges to a unique point called the real image point; light diverging from real image point must then enter an observer's eye for the image point to be observed, or (b) diverges from a different unique point; if part of this light flux then enters an observer's eye, the virtual image point will be simultaneously formed and observed."
>
> (CC4) "Observer's eye as an additional optical instrument: The screen is a convenience for observation of real image points in space."

(CC5) "Light rays: The image formation process can be represented by a light ray diagram; the light ray is a theoretical tool to represent the direction of light propagation."

Previous works by Bendall and colleagues [14], Colin and Viennot [15], Galili and colleagues [13], Goldberg and McDermott [12], Haagen-Schützenhöfer [16], Haagen-Schützenhöfer and Hopf [11,17], Kaltakci-Gurel and colleagues [10], Galili [18], Galili and Hazan [19], Goldberg and colleagues [20], and LaRosa and colleagues [21] have identified several difficulties of students in the field of light, human vision, and lens optics. The concept inventory presented in this work intends to reveal students' difficulties in the field of lens optics, for instance, as a formative or a summative assessment. The concepts of light and vision serve as the basis for understanding image formation by converging lenses and were, therefore, considered in the CLCI, as well. The most relevant core concepts (CC1-CC5 and additional ones identified in the previously mentioned literature) for the topic of image formation and the corresponding learning difficulties are summarized in Table I.

TABLE I. Description of the most relevant core concepts and the corresponding difficulties that students have when learning about image formation by a converging lens.

| Core concepts | Student difficulties (with overarching student difficulties, if existent, printed in bold and facets listed afterwards) |
|---|---|
| **Light properties (LP):** In theory, light rays propagate infinitely | **"Light is resting brightness."** [17] (p. 92) <br> **LP1:** "Light fills space and remains stationary (hence, the sky is bright)." [14,19] (p. 66) <br> "Light is often described as substance-like, a substance that instantaneously fills spaces as soon as a light source is turned on." [17] (p. 92) <br> **LP2:** "Light remains as a glow around a light source (candle, match, bulb, fire)." [14,19] (p. 66) <br> "Light and light source are often perceived as an entity not separable from each other." [17] (p. 92) <br> **LP3:** "Light is comprised of (many or an infinite number of) light rays which fill space." [14,19] (p. 66) [21] <br> "It is assumed […] that light beams […] cannot overlap without interacting with each other." [17] (p. 92) |
| **Light spread (LS):** Light propagates from each point into all directions <br> "Light emanates leaving each point of the light source and propagates, diverging in all spatial directions." [18] (p. 853) | **"Light beams spread out linearly."** [17] (p. 92) <br> **LS1:** "The 'Flashlight model' (for an isolated light source): Light goes out (like a ray or line) only in radial directions." [18] (p. 853) <br> **LS2:** "The 'Flashlight model' (for a compound situation, source + 'consumer'): The light, which is relevant for the |

| Core concepts | Student difficulties (with overarching student difficulties, if existent, printed in bold and facets listed afterwards) |
|---|---|
| | observed image, goes out in a preferential straight-line direction from a point on the source." [18] (p. 853) |
| **Human vision (HV):** An object needs to be illuminated to be visible and light from the object needs to fall into the eye | **"When it's bright, we see something."** [17] (p. 98) <br> **HV1:** "Objects are observed only if they are located in the observer's field of vision and are not blocked." [19] (p. 64) <br> **HV2:** "Light is needed to illuminate eyes, thus enabling their functioning." [14,19] (p. 65) |
| **Refraction of parallel light (RP):** Light is conserved. Parallel light is refracted by the converging lens such that all rays travel through the focal point and further beyond <br> "When the light meets with a lens, the rule of refraction holds true." [10] (p. 4) | **"Light can become more or less"** [17] (p. 94) <br> **RP1:** "The converging lens collects light at the focal point." [17] (p. 94) <br> **RP2:** "Light is multiplied as it passes through the converging lens." [17] (p. 94) <br> **RP3:** "There is more light behind the lens than in front of it." [17] (p. 94) <br> **RP4:** "Light rays [...] simply end at the focal point after passing through the converging lens." [17] (p. 94) |
| **Left-right & top-bottom swap (TB):** The image is an inversion in a point to the object | **"The converging lens inverts the image, therefore it is upside down."** [17] (p. 102) <br> **TB1:** "A lens turns an image upside-down." [13,19] (p. 72) <br> "The impression is created that converging lenses produce images that are merely upside down on the screen, while the left-right reversal is not perceived." [17] (p. 102) |
| **CC1 / Point-to-point imaging (PI):** An object is always imaged point-to-point through a lens <br> "An object is a collection of object points and an image is a collection of image points." [10] (p. 4) | **"Images travel as a whole."** [17] (p. 100) <br> **PI1:** "A converging lens compels light rays to pass through its centre and an inverted image is obtained." [19] (p. 73) <br> **PI2:** "Entire object and entire image [...] (There is little or no awareness of the role of light in these processes.)" [13] (p. 280) <br> "An optical device is responsible for forming the entire image from the whole object." [13] (p. 280) <br> "An image of the object is travelling (in presence of light) to the lens. The lens inverts the entire image." [18] (p. 854) |
| **Correct screen placement (CS):** The image can be observed clearly only in one point <br> "There is a particular image position for a sharp image of an object to be observed on a screen." [10] (p. 4) <br> "When the screen [is] moved from its original position [...], light originating at each object point now produces a whole region of illumination on the screen, rather than a point of illumination. If the screen is very close to its original position, the observer [...] will see a blurry reproduction." [20] (p. 223) | **CS1:** "If the screen moves towards or away from the lens the image becomes bigger or smaller but remains sharp." [13,19] (p. 72) <br> **CS2:** "After the image reaches the screen it is localized on it." [18] (p. 854) |
| **CC4 / Screen removed (SR):** The screen is not necessary for the existence of the image, only for its observability <br> "A screen is a convenience for observation of real image points in space (i.e., aerial image)." [10] (p. 4) | **"No screen no image."** [17] (p. 103) <br> **SR1:** "Surface for image location and observation: A surface is necessary for the observation and location of any image." [13] (p. 280) <br> "Learners often see the function of the screen in 'catching' or materializing images." [17] (p. 103) <br> **SR2:** "Remove screen and look at image. Where is the image located? =>Other locations" [12] (p. 115) |

| Core concepts | Student difficulties (with overarching student difficulties, if existent, printed in bold and facets listed afterwards) |
|---|---|
| | "[Learners think] the presence of a screen is necessary for image formation. Thereby it is unimportant for the formation of the image, where, i.e. at which distance from the lens or in which relation to the object width the screen is placed. From the point of view of many students, the position of the screen ultimately determines only the size of the image." [17] (p. 103) |
| **Diameter of the lens (DL):** Not the diameter, only the focal length (lens' curvature) affects the image | **"The diameter of the lens determines the size of the image."** [17] (p. 103) |
| **CC2 / Partly covered lens (PC):** As the object is imaged point-to-point, covering the lens partly does not affect the image (except for making it less bright) | **"A half lens produces a half image."** [12,19] (p. 72) <br> **PC1:** "A half-lens produces a half-image. The rest of the image […] is blocked." [19] (p. 61) <br> **PC2:** "When half of a lens is covered, half the light rays from the object are blocked, so only half of its image appears." [19] (p. 73) |
| **CC5 / Object larger than lens (OL):** Object size compared to lens diameter does not affect the imaging of the object <br> "The light ray is a representational tool to show the direction of light prorogation. The special rays [i.e., construction rays] serve as an algorithm to locate the position of an image. […] However, special rays are sufficient but not necessary in order to form an image point." [10] (p. 4) | **OL1:** "A lens can only image objects completely if they are not larger than the lens itself." [17] (p. 103) <br> **OL2:** "If the object is larger than the lens diameter, the construction [of the parallel ray through the lens] fails and learners conclude that no image can be formed." [17] (p. 103) |
| **Rule of motion (RM):** If the object is far away, the image is closer to the focal point on the other side; if it is close to the focal length, the image is far away on the other side; an object in the focal point cannot be imaged by the lens (this rule can be directly derived from the three rules of refraction) | No student difficulty was specifically identified for the understanding of the rule of motion in previous literature. <br> We suggest as distractor: Imaging works like mirroring: object and image of the object will always have the same distance to the lens. |
| **CC3 / Virtual image (VI):** If the object is closer to the lens than one focal length, there is a virtual image on the same side and upright (this can be directly derived from the three rules of refraction) | No student difficulty was specifically identified for the understanding of virtual image construction in previous literature. <br> We suggest as distractor: The image of the object will always appear on the opposite side of the lens. |

## C. Previously developed test instruments

While tests exist within the context of geometrical optics, they have certain limitations: First, they often do not address the topic of image formation by a converging lens at all (e.g., [22–25]) or not comprehensively, considering all the well-known student difficulties that are outlined in Table I (e.g., [10,19,26,27]). Second, these instruments were partly constructed for postsecondary rather than secondary students (e.g., [10,27]) and often lack differentiate empirical validation.

Therefore, our newly developed CLCI fills the existing gap of comprehensive test instruments for middle school lens optics by presenting an assessment of students' conceptual understanding of image formation by a converging lens.

### III. INSTRUMENT DEVELOPMENT, CONTENT, AND FORMAT

#### A. Instrument development

We gathered a team of experts in physics, physics education, and educational psychology, consisting of university faculties, university research and teaching staff, and middle-school teachers to jointly develop the test. Two researchers took the lead in the item development, the other team members were frequently consulted and involved throughout the whole development process. First, frequently observed difficulties of students concerning converging lenses were identified in the literature (see above). Second, items from existing optics tests were screened as to whether they covered any of the identified students' difficulties; then, they were judged concerning their potential to be modified and used in this test. Third, we agreed upon the format of the test, which is described in detail in the section "Format," below. Then, we generated multiple choice test items based on the students' difficulties; each incorrect answer option (distractor) was constructed such that it represented a specific facet of a student's difficulty. Each of the answer options was developed with text and a picture to increase the clarity and understandability of what was meant by this answer. The text passages were written in easy language so that especially middle school students can understand their meaning. The pictures were designed in simple and meaningful sketches to focus on the essential.

#### B. Content

The complete test instrument in its original format containing 16 items can be found in Appendix A. The content of the items is based on the core concepts and student difficulties that were summarized in Table I. Table II describes the concrete link of each item to the core concepts and the student difficulties.

TABLE II Description of the single items and the core concepts and student difficulties they are based on ("Basis"). Answer options of each item are labeled A, B, C, etc. The correct answer option/s is/are marked with an asterisk.

| Item | Short description of the answer options within each item | Basis |
|---|---|---|
| Q1 | A: Light behaves similarly to a fluid | LP1 |
|  | B: Light is inside a light source | LP2 |
|  | C*: Light propagates infinitely | LP |
|  | D: Light beams can collide and bounce off each other | LP3 |
| Q2[a] | A: All light rays propagate away from an imaginary center of the light source | LS1 |
|  | B: All light rays propagate in the same direction from the surface of a light source | LS2 |
|  | C*: Light propagates from each point into all directions | LS |
| Q3 | A: Eye and object are both in the shadow, but there is no shielding between eye and object | HV1 |
|  | B: Only the object is in the shadow, but the eye is illuminated | HV2 |
|  | C*: Light can spread from the source to the object and into the eye (version 1) | HV |
|  | D*: Light can spread from the source to the object and into the eye (version 2) | HV |
| Q4 | A: Light rays end in the focal point and continue like a laser | RP1, RP |
|  | B: Behind the lens, there are more light rays than in front of the lens | RP2 |
|  | C: The lens does not affect the light rays' path but makes them brighter | RP3, RP |
|  | D*: Light is conserved and all rays travel through the focal point and further beyond | RP |
|  | E: Light rays are brighter behind the lens and stop in the focal point | RP3, RP4 |
|  | F: Light rays are multiplied and continue on their parallel path after the focal point | RP2, RP |
| Q5 | A: The lens does not invert the image at all | TB |
|  | B*: The image is an inversion in a point to the object | TB |
|  | C: Only left and right are inverted | TB1 |
|  | D: Only top and bottom are inverted | TB1 |
| Q6 | A: Only left and right are inverted | TB1 |
|  | B: The lens does not invert the image at all | TB |
|  | C: Only top and bottom are inverted | TB1 |
|  | D*: The image is an inversion in a point to the object | TB |
| Q7 | A: The light rays travel through the lens without crossing | PI1 |
|  | B: The lens plane flips the object when traveling through the lens | PI2 |
|  | C: The center of the lens flips the object when traveling through the lens | PI1, PI2 |
|  | D*: An object is always imaged point-to-point through a lens | PI |
| Q8 | A: The image is smaller because it is further away | CS1 |
|  | B: The image is the same size but blurred because of the distance | CS2, CS |
|  | C*: The image is larger and blurred | CS |
|  | D: It does not matter where the screen is placed, the image will always be the same | CS2 |
| Q9 | A: The screen needs to catch the image | SR1, SR2 |
|  | B*: The image will be at the same place, only not observable | SR |
|  | C: No screen causes the light rays to not pass the lens | SR2 |
|  | D: No screen leads to no rotation of the image | SR1, SR2 |
|  | E: All rays will be collected in the focal point like with parallel rays | SR1 |

| Item | Short description of the answer options within each item | Basis |
|---|---|---|
| | F: No screen leads to an image at infinity | SR1, SR2 |
| Q10 | A*: The image remains the same | DL |
| | B: If the diameter of the lens is larger, the image becomes larger | DL |
| | C: If the diameter of the lens is larger, the image becomes smaller | DL |
| Q11 | A*: The image remains the same | DL |
| | B: If the diameter of the lens is smaller, the image becomes larger | DL |
| | C: If the diameter of the lens is smaller, the image becomes smaller | DL |
| Q12 | A: The image of the object does not fit through the lens, leading to no image at all | PC1 |
| | B*: Covering the lens partly does not affect the image except for making it less bright | PC |
| | C: Only the part of the object that fits through the pinhole can be imaged on the screen | PC1 |
| | D: There is no difference in the image with pinhole | PC |
| | E: Only the outer parts of the image are cut off | PC1 |
| Q13 | A: The parts of the object that do not fit through the lens are cut off | OL1 |
| | B: The image of the object does not fit through the lens, leading to no image | OL2 |
| | C: If the object is larger than the lens, a smaller image of the object is created | OL |
| | D*: Covering the lens partly does not affect the image except for making it less bright | OL |
| Q14 | A: Only the part of the object that fits over the cover can be imaged on the screen | PC2 |
| | B: As not all constructing lines travel through the lens, there will be no image at all | PC2 |
| | C*: Covering the lens partly does not affect the image except for making it less bright | PC |
| | D: The parts of the object that fit over and under the cover can be imaged on the screen | PC2 |
| Q15 | A: An object in the focal point of the lens creates an image in the other focal point | RM |
| | B*: An object in the focal point cannot be imaged by the lens | RM |
| | C*: An object far away from the lens creates an image close to the lens | RM |
| | D: An object far away from the lens creates an image far away from the lens | RM |
| | E: An object close to the focal point creates an image close to the other focal point | RM |
| | F*: An object close to the focal point creates an image far away from the lens | RM |
| Q16 | A: The image is upside-down on the other side of the lens | VI |
| | B: The image is upside-down on the same side of the lens | VI |
| | C: The image is upright on the other side of the lens | VI |
| | D*: The image is upright on the same side of the lens | VI |

[a]Item Q2 was removed from the final version of the CLCI.

## C. Format

When developing the CI, we paid special attention to a practical format of test implementation and test score evaluation. The goal was to create a test that could be used easily in schools and for empirical research alike. Therefore, it should be as parsimonious as possible, reflect a broad range of difficulties, and its scoring should be fast and objective. Still, the test should cover all the relevant core concepts to provide a comprehensive view on students' conceptual understanding of the topic. We decided for a one-tier test to keep the test structure simple and save time during test implementation, and for a multiple-choice format to ensure the quick and objective evaluation of students' answers.

Of the 16 items, 11 items conformed to the default format, that is, with text and picture both in the item stem and in all answer options (items Q1-Q4 and Q6 are different). One exemplary item in the default format (item Q7) is shown in Fig. 3.

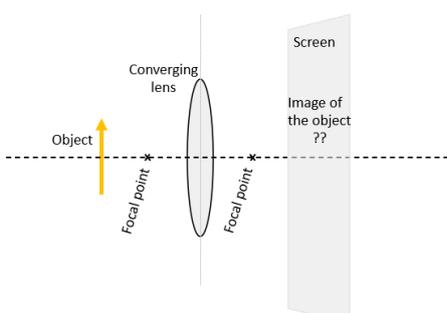

FIG. 3. Exemplary item of the CLCI (item Q7) with the default format: text and picture in both item stem and all answer options.

The combination of text and picture in both the item stem and the answer options is a strength of the CLCI, because this format makes it more independent from purely schematic representations of the rays. This can be especially important for identifying the prior difficulties that students have before they are taught about converging lenses in school, because, at this point in time, they are mostly not familiar with the schematic representation including rays. The items had either 3 (3 items), 4 (9 items), 5 (1 item), or 6 (3 items) answer options. Table B.I in Appendix B gives a detailed overview of the formats of all test items and their specificities. Prior to solving the CLCI, students were given oral or written test instructions and a written introduction to the test, including an example item. Both the test instructions and the introduction can be found in Appendix A. The instructions and the introduction informed test takers that each item could, in principle, have multiple correct answers, which should encourage students to actually read through all answer options in detail before choosing their answer. The minimum number of correct answer options was 1 and the maximum number was 3. We used a partial credit model to score the test: Full credit, that is, 2 points were given, if a student had selected all the correct answers and no wrong answers. Partial credit, that is, 1 point was given, if a student had selected only some of the correct answers or if a student had selected the correct answer(s) plus incorrect ones. No credit was given if a student had selected only the wrong answer(s). This scoring rule was illustrated in the example item as part of the test introduction. Notably, the majority of the items had only one correct answer (14 items); in addition, there was one item with two correct answer options out of four (Q3), and one item with three correct answer options out of 6 (Q15). This aspect is discussed in the limitations and outlook sections of this article.

## IV.  QUANTITATIVE VALIDATION

In this chapter, we first report analyses of the test with a special focus on the single items, beginning with an analysis of the distractors we used in the test. We further conducted the analyses suggested by Jorion and colleagues [9] and evaluated the test corresponding to their "categorical judgement scheme and assignment rules for evaluating a CI" (see [9], p. 482). The suggested analyses include item and test analyses according to classical test theory and item analyses according to item response theory [9].

As these first analyses all indicated that item Q2 should be removed from the test, we followed the suggestion of Jorion and colleagues [9] and excluded students' answers for item Q2 from the subsequent analyses conducted with the test in total. These analyses include structural analyses of the test (i.e., exploratory factor analysis (EFA) and confirmatory factor analysis (CFA), as suggested by [9]), the distribution of students' test scores in pre- and posttest, the correlation of students' grades and their test scores, and the test-retest reliability we evaluated from pre- and posttests as well as from post- and follow-up tests. Table III gives an overview of the analyses conducted within the quantitative validation of the CLCI.

TABLE III. Summary of analyses conducted for quantitative test analyses of the CLCI.

| Analysis | Measures |
| --- | --- |
| Distractor analysis | Percentage of students selecting each answer option in pretest and posttest |
| Classical test theory | |
|     Item statistics | Item difficulty |
| | Item discrimination |
|     Total score reliability | Cronbach's alpha of total score |
| | Cronbach's alpha-with-item-deleted |
| Item response theory | Individual item measures |
| | Fit of the items with the model |
| Structural analyses | |
|     Exploratory factor analysis | Fit of the model to the data |
|     Confirmatory factor analysis | Item loading |
| | Comparative fit index |
| | Root-mean-square error approximation |
| Distribution of students' test scores | Increase of students' test scores from pre to post (t-test) |
| Correlation of students' grades and test scores | Pearson correlation of the grades with pre, post-, and follow-up tests |
| Test-retest reliability | Intraclass Correlation Coefficient of pre- and posttest |
| | Intraclass Correlation Coefficient of post- and follow-up test |

All analyses were conducted using R (version 4.1.1) and the package *psych* (version 2.1.9) to calculate Cronbach's alpha values, the package *mirt* (version 1.34.11) for the item response theory analyses, the packages *psych* and *GPArotation* (version 2014.11-1) to calculate the EFA, and *lavaan* (version 0.6-9) for the CFA.

## A. Method of quantitative validation

### 1. Test administration

We administered the CLCI to students three times: first, approximately one week before students were introduced to converging lenses in their school lesson (as a pretest), second, one day after their lessons on converging lenses in school (as a posttest), and third, approximately 4 to 8 weeks after their lessons on converging lenses in school (as a follow-up test). Data was gathered either in an in-person setting in school where the test administrators came into the class and administered the CLCI in a standardized procedure or during COVID-19 induced school closures, during which the CLCI was administered online using the software Qualtrics. In each situation, students answered the test individually. All instructions that were given by the in-person instructors were also provided in written format in the online survey. Before solving the CLCI, students went through an example item unrelated to the topic of the test to understand how the items were organized. Students were asked to solve the CLCI alone and without the help of the internet, learning material, or other persons such as classmates, siblings, or parents. Both in the in-class scenario and in the online setting, most students took around 20 minutes to solve the CLCI. In the in-class scenario, the time for solving the test was restricted to a maximum of 30 minutes; in the online setting, the time was not restricted. However, there was no participant in the online setting who took longer than 30 minutes to solve the CLCI. Participants who answered fewer than 14 items of the 16-item CLCI were excluded from the sample. The whole instrument contained the CLCI and demographic questions about students' ages, genders, mother tongues, and physics grades. Students were asked about their past two science/physics grades and their grade on the last physics exam (all on a scale of 1 = *very good* to 6 = *insufficient*); their average physics grades were then calculated as the means of these three grades.

For the CLCI quantitative validation analyses, we used the pre-, post-, and follow-up test data. The reason for administering the test to students at three time points was that we wanted to show how students' preconceptions related to converging lenses eventually changed after they were taught about converging lenses in school (pre to post) and we wanted to measure the test-retest reliability of this test instrument from pre to post and after a few weeks (post to follow-up test). We claim that our test instrument can detect conceptions that students possess prior to optics classes (pretest) and that it can measure how elaborated students' conceptual understanding may become after optics classes (posttest). Thus, the test instrument should be able to reflect changes in students' understanding, as indicated by a medium-sized retest reliability of pretest and posttest. At the same time, once conceptual understanding has been obtained, it should be relatively stable, as reflected in a high test-retest reliability of posttest and follow-up test. We used only the posttest data for the item and test analyses according to classical test theory, item response theory, and for structural analyses of the test, as this is the standard procedure when CIs are validated [9]. In the posttest, students' conceptions regarding the topic are more coherent and less influenced by potential guessing behavior and, hence, provide more information regarding the quality of the items and the test as a whole.

## *2. Sample*

In total, $N = 349$ students took the pretest and $N = 318$ students took the posttest. These samples are overlapping, in that 277 students took both tests, whereas the remaining 113 students missed either the pretest (41 students) or the posttest (72 students). $N = 72$ students who took the posttest also participated in the follow-up test. Students from grade 7 (and ten students from grade 9) of German Gymnasiums and middle schools participated in the posttest sample, aged $M_{age} = 12.60$ years and $SD_{age} = 0.68$ years. 46.23% of these students were male, 43.08% were female, and 10.69% did not indicate their gender. Most students had German as their mother tongue (88.42% in the posttest). Students participating in the posttest had an average physics grade of $M_{grade} = 2.39$ and $SD_{grade} = 0.77$ on the German grade scale of 1 (very good) to 6 (insufficient). Students' age, gender, mother tongue, and physics grade in the pretest and follow-up test samples were not significantly different from those of the posttest sample; all t-test comparisons were not significant, with *p*-values > .16.

## **B.  Distractor analysis**

For the distractor analysis, we evaluated which answer options students chose in pre-, post-, and follow-up test. Students were allowed to select multiple answer options for each item. Figure 4 displays the percentage of the students selected with answer options for all 16 items of the pre-, post-, and follow-up test. Correct answers are marked with an asterisk in Fig. 4. Table C.I in Appendix C shows all values for the percentages of students selecting the item answer options for pre-, post-, and follow-up tests.

In the pretest, students selected the distractors as well as the correct answer options on a frequent basis. All answer options were chosen by at least 5% of the students in the sample, except for item Q2, where answer option B was chosen by 2% of the students only. In items Q1, Q3, Q4, Q5, Q6, Q7, and Q8, the correct answer option(s) was/were selected by most students already in the pretest, suggesting that these items were relatively easy. However, in these items, the distractors were still chosen frequently.

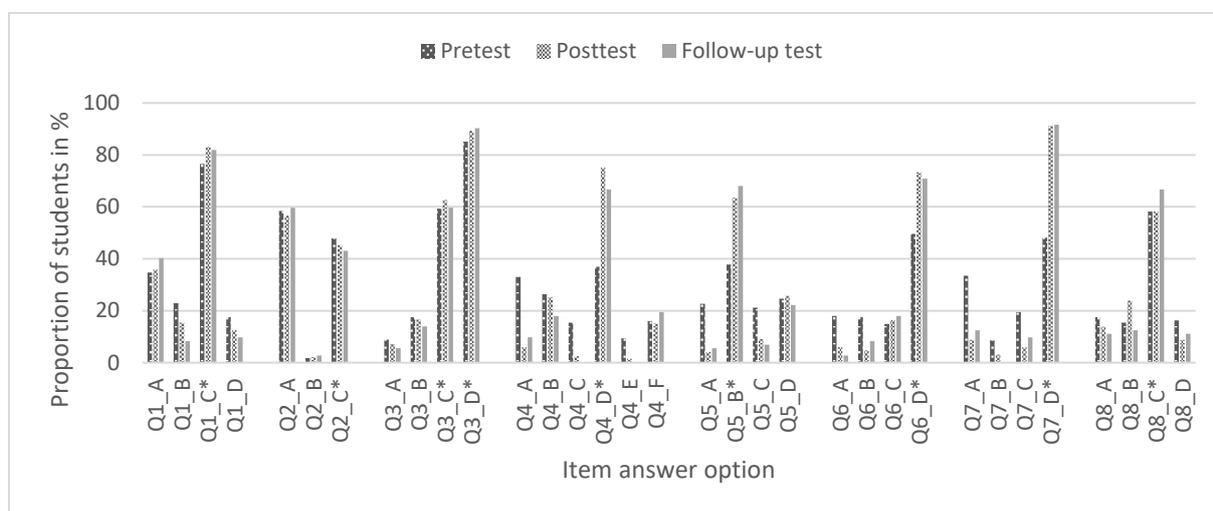

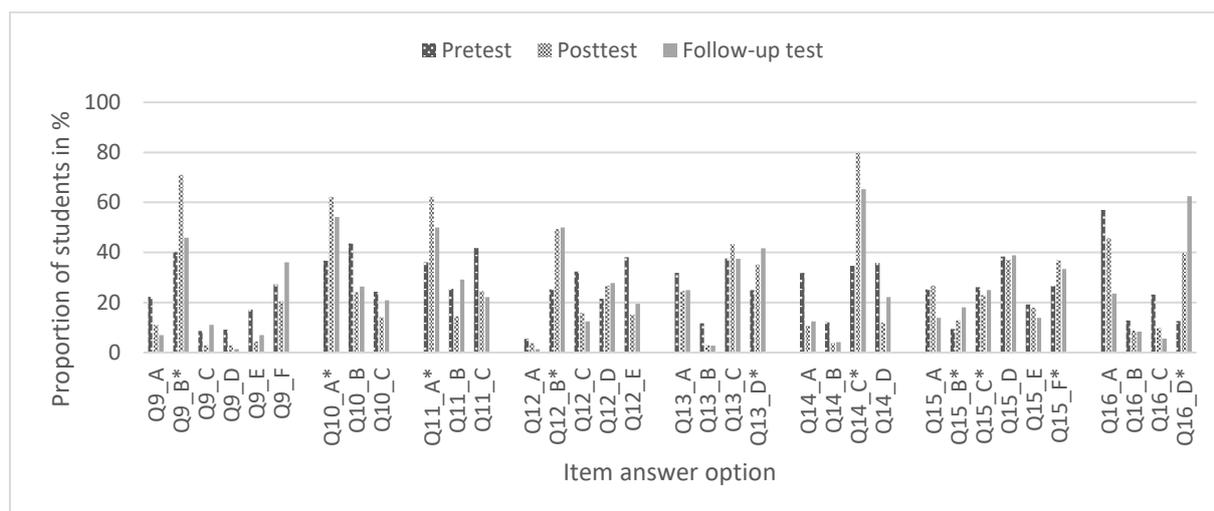

FIG. 4. Percentage of the students who selected each answer option for all 16 items in the pre-, post-, and follow-up test (split up in Q1-Q8 and Q9-Q16). Correct answer options are indicated by an asterisk.

When comparing the pre-, post- and follow-up test in Fig. 4, it is apparent that students chose the correct answer options in the posttest and follow-up test more frequently than in the pretest. Moreover, in the posttest and the follow-up test, the distractors in most items were chosen much less frequently than in the pretest. Most students selected the correct answer option(s) in the posttest for all items except for the items Q2, Q13, Q15, and Q16. In the follow-up test, most students selected the correct answer option(s) in all items except for item Q2 and item Q15, in which the correct answer option Q15_C was still chosen less frequently than the incorrect answer option Q15_D.

When looking in detail at item Q2, it appears that students did not change their minds about this item from the pre- to post- and follow-up test, and that more students chose the incorrect answer option A than the correct answer option C at all measurement points. Moreover, only 2-3% of the students selected distractor B in all three measurement points, which is the smallest percentage of students who selected an answer option in the whole test. As it seems to be evident to most students that distractor B is not a correct answer here, students are only left with two other answer options and they could simply guess which of the two remaining options, A or C, might be correct. This indicates item Q2 is not a good functioning item in this test.

### C. Analyses according to classical test theory

All values corresponding to the item statistics and the total score reliability can be found in Table C.II in Appendix C.

#### *1. Item statistics*

We calculated the difficulty of each item by evaluating what ratio of students solved the item correctly (probability of solving the item correctly, on a scale of 0 = *the item was solved incorrectly* to 1 = *the item was solved correctly*).

We calculated the item difficulties for all original 16 items of the CLCI. They ranged from .28 (Q15, difficult) to .78 (Q14, not so difficult), except for item Q7, which had a difficulty of .87 and was therefore rather easy (see Fig. 5). According to the scheme of Jorion and colleagues [9] (p. 482), this indicates the difficulty is "good" (almost "excellent"). According to Ding and Beichner [28], who suggest a desired value of the difficulty index between .3 and .9, all our items lie within the desired range.

Item discrimination was calculated by considering the correlation between the points reached in one item and the sum of points reached in all other items of the test, except for the item of interest across the 318 students.

We calculated the item discrimination for all original 16 items, resulting in values between .18 (Q1) and .45 (Q5), except for item Q2, which had a discrimination of –.03 (see Fig. 5). Figure 5 shows the items' discriminations against the items' difficulties for all 16 original CLCI items in one plot. The fact that Q2's item discrimination is negative was another argument to remove item Q2 from the CLCI.

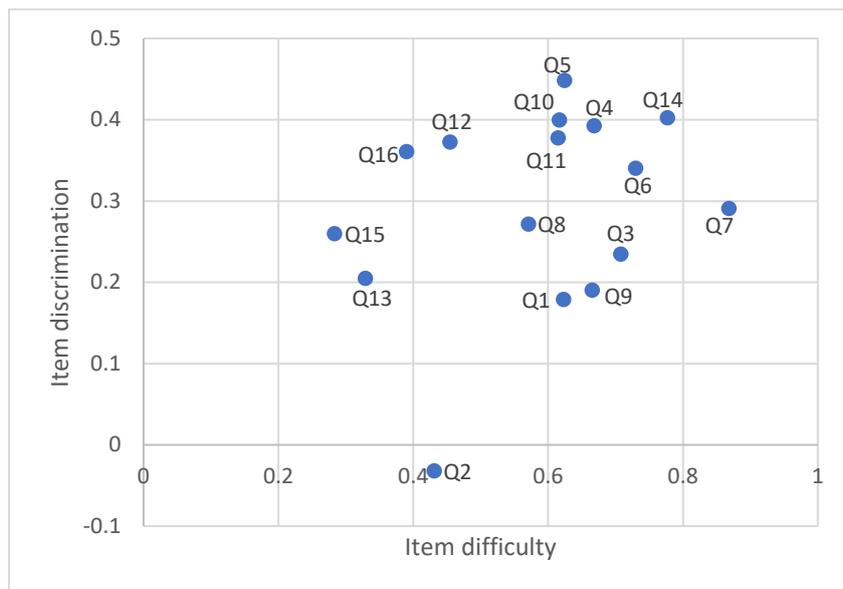

FIG. 5. Item statistics for all original 16 items of the CLCI. Item Q2 shows negative discrimination.

As a consequence, we calculated the items' discrimination once again after item Q2 was removed from the test. At that point, only one item showed a discrimination below .20, namely, Q1 with .18. All the other items ranged from .20 (Q9) to .44 (Q5; see Table C.II in Appendix C). According to Jorion and colleagues [9] (p. 482), the item discrimination can therefore be considered "good" (almost "excellent"). Ding and Beichner [28] suggest a stricter discrimination index of equal to or higher than .3, which is the case for 8 out of the 15 CLCI-items.

### 2. Total score reliability

Even though the test was constructed based on multiple conceptual ideas and, therefore, is likely to be multidimensional, we also computed Cronbach's alpha for the whole test instrument. Cronbach's alpha for the subscales identified in the structural analyses section are considered later.

Cronbach's alpha of the total score for the 16-item CLCI was .69. Cronbach's alpha-with-item-deleted for the 16-item CLCI showed values between .66 (when deleting Q5 or Q10) and .69 (when deleting Q1, Q9, or Q13), except for item Q2, which showed a higher alpha-with-item-deleted of .72 (see Table C.II in Appendix C). Again, item Q2 being an outlier suggests removing this item from the test.

Hence, we calculated Cronbach's alpha for the 15-item CLCI (.72) and Cronbach's alpha-with-item-deleted for the remaining 15 items again. The values for Cronbach's alpha-with-item-deleted now ranged from .69 (when deleting Q4, Q5, or Q10) to .72 (when deleting Q1, Q9, or Q13; see Table C.II in Appendix C). However, deleting Q1, Q9, or Q13 would not lead to further improvement of Cronbach's alpha compared to the 15-item CLCI Cronbach's alpha of .72. After removing item Q2, according to Jorion and colleagues [9] (p. 482), the Cronbach's alpha of the total score (.72) can be considered "average" and the Cronbach's alpha-with-item-deleted "excellent."

### D. Analyses according to item response theory

Jorion and colleagues [9] additionally suggest item response theory analyses with all original items of the evaluated CI. We performed an analysis of the CLCI with the package *mirt* in R, using a unidimensional IRT model. Since students could receive 0, 1, or 2 points for each item, a dichotomous model was not possible, so we used a generalized partial credit model. We assessed the adequacy of model fit using the root mean square error of approximation (RMSEA), the standardized root mean square residual (SRMSR), and the comparative fit index (CFI). The obtained RMSEA value = .081 (95% CI [.071, .091]) suggests that this model has a rather fair to poor fit (recommended cutoff value RMSEA <= .06); however, the SRMSR value = .079 is below the recommended cutoff value of SRMSR <= .08 and the CFI = .763 is below the recommended .95 threshold [29]. This indicates an overall acceptable fit of the generalized partial credit model to the data.

Next, we assessed how well each item fits the model using the S-$X^2$ item fit index for polytomous IRT models [30]. The RMSEA values of the S-$X^2$ item index fit were below the suggested threshold of .06 for all items (cf. Tab. 4), indicating an adequate fit of all items to the model. According to Jorion and colleagues [9] (p. 482), the individual item measures according to item response theory can therefore be considered "excellent."

Next, we computed the item parameters for the model. The parameters a, b1, and b2 for all 16 items are shown in Table IV.

TABLE IV. RMSEA values of the S-$X^2$ item index fit and item parameters for the IRT model.

| Item | RMSEA of S-$X^2$ | a | b1 | b2 |
|---|---|---|---|---|
| Q1 | 0.027 | 0.282 | -3.369 | 0.066 |
| Q2 | 0.040 | -0.025 | -100.118 | 88.527 |
| Q3 | 0.028 | 0.460 | -3.949 | -0.327 |
| Q4 | 0.000 | 0.739 | 0.064 | -1.627 |
| Q5 | 0.000 | 0.793 | 3.110 | -4.083 |
| Q6 | 0.027 | 0.603 | 5.739 | -7.864 |
| Q7 | 0.037 | 0.589 | -0.624 | -3.931 |
| Q8 | 0.040 | 0.363 | 7.892 | -8.798 |
| Q9 | 0.045 | 0.288 | 3.780 | -6.608 |
| Q10 | 0.000 | 0.718 | 5.310 | -6.254 |
| Q11 | 0.035 | 0.681 | 5.043 | -6.003 |
| Q12 | 0.013 | 0.612 | 2.836 | -2.444 |
| Q13 | 0.039 | 0.270 | 10.854 | -7.932 |
| Q14 | 0.040 | 0.800 | 1.740 | -4.055 |
| Q15 | 0.000 | 0.583 | -0.074 | 4.748 |
| Q16 | 0.032 | 0.562 | 6.753 | -5.762 |

The values of the slope parameter (a-parameter) ranged from -0.025 (Q2) to 0.800 (Q14). Larger values (steeper slopes) indicate better differentiation of learners' abilities. Item Q14 was, therefore, the most discriminating item, with a slope estimate of 0.800; item Q2 shows a problematic negative slope. The problematic item parameters for item Q2 can also be observed in the diagram of the probability function of item Q2 in Fig. 6. The probability functions for all 16 items are depicted in Appendix C. The blue curve represents the probability of scoring 0 points in this item, the pink curve represents the probability of scoring 1 point in this item, and the green curve corresponds to scoring 2 points in this item. The totals of all three probability values for a certain learner's level always add up to 1.0.

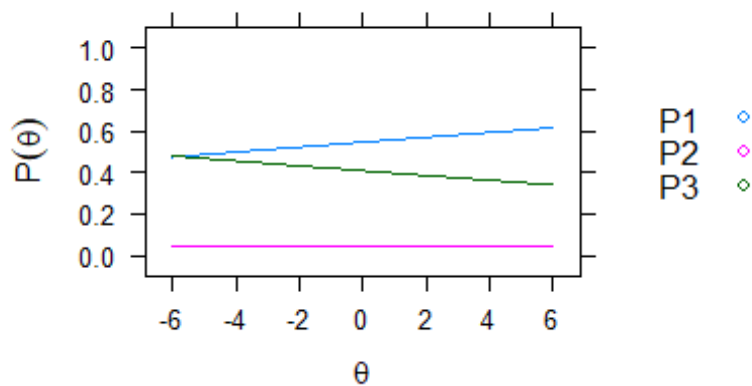

FIG. 6. Probability function of item Q2. P1 (blue) indicates 0 points, P2 (pink) 1 point, and P3 (green) 2 points scored. The axes represent the learner's ability (x-axis, theta) and the probability of solving this item correctly (y-axis).

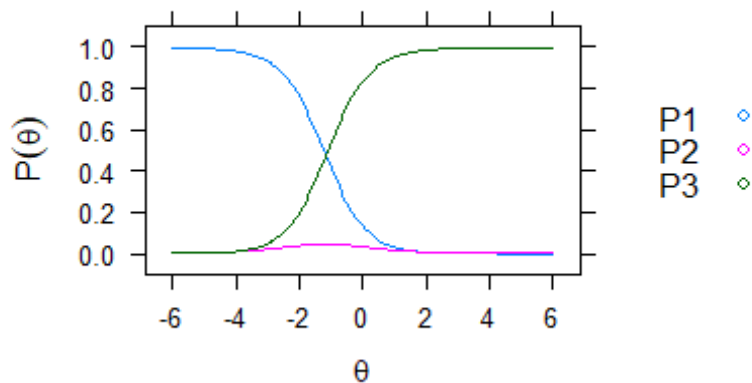

FIG. 7. Probability function of item Q14. P1 (blue) indicates 0 points, P2 (pink) 1 point and P3 (green) 2 points scored. The axes represent the learner's ability (x-axis, theta) and the probability of solving this item correctly (y-axis).

Figure 7 shows the probability function of item Q14. For this item, learners with low ability mostly scored 0 points, learners with medium ability scored 0, 1, or 2 points, and learners with high ability mostly score 2 points. This indicates an adequate solving pattern for this item. Item Q2 does not show this adequate solving pattern; learners with low ability scored equally 0 or 2 points in this item, learners with medium ability scored mostly 0 points, second most frequently they scored 2 points, learners with high ability similarly scored mostly 0 points and only second most frequently they scored 2 points. This again supports the idea to remove item Q2 from the test.

### E. Structural analyses of the test

For the structural analysis of the test, we excluded item Q2, as it was clear from the prior analyses that this item would ultimately be removed from the test ("Any items with poor properties should be removed from subsequent structural analyses." [9], p. 457). As suggested by Jorion and colleagues [9], we first conducted an exploratory factor analysis and then a confirmatory factor analysis: "If the exploratory factor analysis indicates recoverable structure, either comparable to or different from the developer's originally designated constructs, confirmatory factor analyses […] can also be conducted to test specific hypotheses about structure." [9] (p. 457). Thereby, we followed the approach recommended by Anderson and Gerbing [31] (p. 421): "Ideally, a researcher would want to split a sample, using one half to develop a model and the other half to validate the solution obtained from the first half."

#### *1. Exploratory factor analysis*

We calculated the exploratory factor analysis with one half of our posttest dataset of the CLCI ($n = 159$). First, we conducted a parallel analysis (factor method: minimum residual solution, eigen values: principal axis factor analysis) and identified from the Scree plot (attached in Appendix C) that a number of factors between 4 and 8 would be reasonable; four factors were suggested by R. Therefore, we computed the exploratory factor analysis for all five scenarios, using oblique rotation, as suggested by Jorion and colleagues [9], due to possible underlying relations between the single factors. We compared the models of 4 to 8 factors concerning their root mean square of the residuals (RMSR) value, their RMSEA value, and their Tucker-Lewis-Index (all model values are summarized in Table C.III in Appendix C). The model with 8 factors showed the best fit to the dataset, considering the model values and "compared to the developers' concept groupings" [9] (p. 488).

Figure 8 shows the distribution of items to the 8 factors. Item Q16 was not single-loading and did not load on one factor with more than 0.3: it loaded with 0.27 on MR6 and with 0.25 on MR5. Because Fig. 8 uses a factor loading threshold of 0.3, Q16 is depicted without being assigned to a factor.

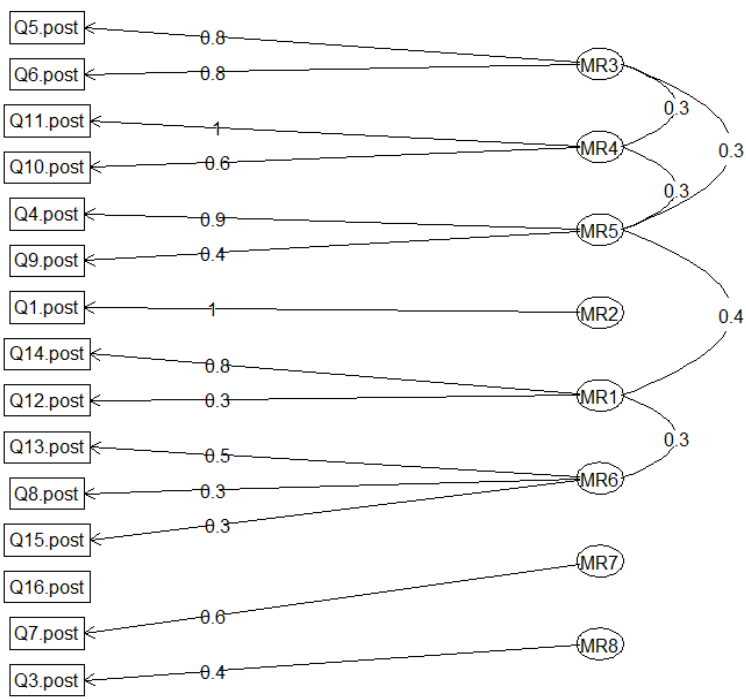

FIG. 8. Distribution of the 15 CLCI items to the 8 factors MR1-MR8 determined in an exploratory factor analysis.

The 8 factors can be matched with the subscales as indicated in Table V. Factor MR3 (containing Q5 and Q6) refers to the core concept TB. Factor MR4 (containing Q10 and Q11) refers to the core concept DL. Factor MR5 (containing Q4, Q9, and possibly Q16) refers to the core concepts RP, SR, and possibly VI. These items have in common that there is no screen visualized; thus, they might rely on a common underlying erroneous conception about the image formation by a converging lens without having a projection surface. Factor MR2, which only contains Q1, refers to the core concept LP. Factor MR1 (containing Q12 and Q14) refers to the core concept PC. Factor MR6 (containing Q8, Q13, Q15, and possibly Q16) refers to the core concepts CS, OL, and RM. All these core concepts might be related to the common student's misconception that imaging and mirroring are mixed up. The core concept underlying item Q16 (VI) would also fit this broader idea. Factor MR7 contains only Q7, referring to the core concept PI. Last, factor MR8 (containing only Q3) belongs to the core concept HV.

All in all, the model chosen based on the exploratory factor analysis also conforms to the predicted constructs and can thus be evaluated as "excellent" according to Jorion and colleagues [9] (p. 482).

### 2. *Confirmatory factor analysis*

To further confirm these factors, we calculated a confirmatory factor analysis (CFA) with the other half of the posttest dataset ($n = 159$). We used the factors determined in the EFA; item Q16 was added to MR6 because it showed a higher load to this factor than to MR5. The results of the CFA are summarized in Table V in the columns "factor loadings" and "estimated error variance."

TABLE V. Cronbach's alpha values for the subscales containing more than one item.

| Subscale | Core concepts | Items | Factor loadings | Estimated error variance | Cronbach's alpha |
|---|---|---|---|---|---|
| MR1 | PC | Q12 | 0.58 | 0.66 | .46 |
|  |  | Q14 | 0.43 | 0.82 |  |
| MR2 | LP | Q1 | 1.00 | 0.00 | n/a |
| MR3 | TB | Q5 | 1.87 | -2.49 | .78 |
|  |  | Q6 | 0.35 | 0.88 |  |
| MR4 | DL | Q10 | 0.85 | 0.28 | .80 |
|  |  | Q11 | 0.83 | 0.31 |  |
| MR5 | RP, SR | Q4 | 0.58 | 0.67 | .47 |
|  |  | Q9 | 0.42 | 0.82 |  |

| MR6 | CS, OL, RM, VI | Q8  | 0.40 | 0.84 | .41 |
|-----|----------------|-----|------|------|-----|
|     |                | Q13 | 0.12 | 0.99 |     |
|     |                | Q15 | 0.36 | 0.87 |     |
|     |                | Q16 | 0.40 | 0.84 |     |
| MR7 | PI             | Q7  | 1.00 | 0.00 | n/a |
| MR8 | HV             | Q3  | 1.00 | 0.00 | n/a |

The factor loading of Q5 to MR3 is larger than 1, with a negative estimated error variance. This can probably be explained by the fact that Q5 and Q6 are too similar in students' answering behavior. Deleting either item Q5 or item Q6 from the model resolves this problem. However, as all the other analyses we conducted suggested that item Q5 and item Q6 are unproblematic, and especially that students' interviews suggested that the two items are indeed different from each other in the way they were understood by students, we decided to keep both items.

All item loadings can be confirmed to be valid with the other half of the data set and the CFA, except the loading of Q13 to MR6. All items load onto the respective factors with loadings higher than .3; only Q13 shows an item loading of .12. Therefore, the item loading of the CFA can be considered "good" according to Jorion and colleagues [9] (p. 482).

The comparative fit index of the model is higher than .9 with a value of 1.000 and the root-mean-square error approximation of the model is lower than .03 with a value of .000; both are, therefore, "excellent," according to Jorion and colleagues [9] (p. 482).

Furthermore, we calculated the internal Cronbach's alpha values for each of the subscales containing more than one item (Table V, column "Cronbach's alpha").

These Cronbach's alpha values are partly very high, showing good internal consistency, and partly relatively low, which is normal, considering that these scales consist of few items and that the items in MR5, MR1, and MR6 do not duplicate, but rather, complement each other and serve the same underlying subscale.

## F. Distribution of students' test scores in pre- and posttest

For an analysis of the distribution of students' total test scores in pre- and posttest, we added up all the item scores students gained in the test, except the score for item Q2. With a maximum of 2 points per item, students could achieve a maximum of 30 points in the whole test. We plotted a histogram of students' total test scores for pre- and posttest in Fig. 9.

Figure 9 shows that students' total scores shifted from the lower half of the scale to the upper half of the scale from pre- to posttest and that both distributions correspond to approximately normal distributions. These findings were tested statistically, too: The test scores' mean and standard deviation shifts from $M_{pre}$ = 11.07 points out of 30 (36.89%), $SD_{pre}$ = 4.37 (14.55%) to $M_{post}$ = 17.84 points out of 30 (59.47%), $SD_{post}$ = 5.80 (19.34%). This shift is statistically highly significant, indicated by a t-test ($p < .0001$). Both the pretest data ($p < .0001$) and the posttest data ($p = .001$) are normally distributed, indicated by Shapiro-Wilk normality tests.

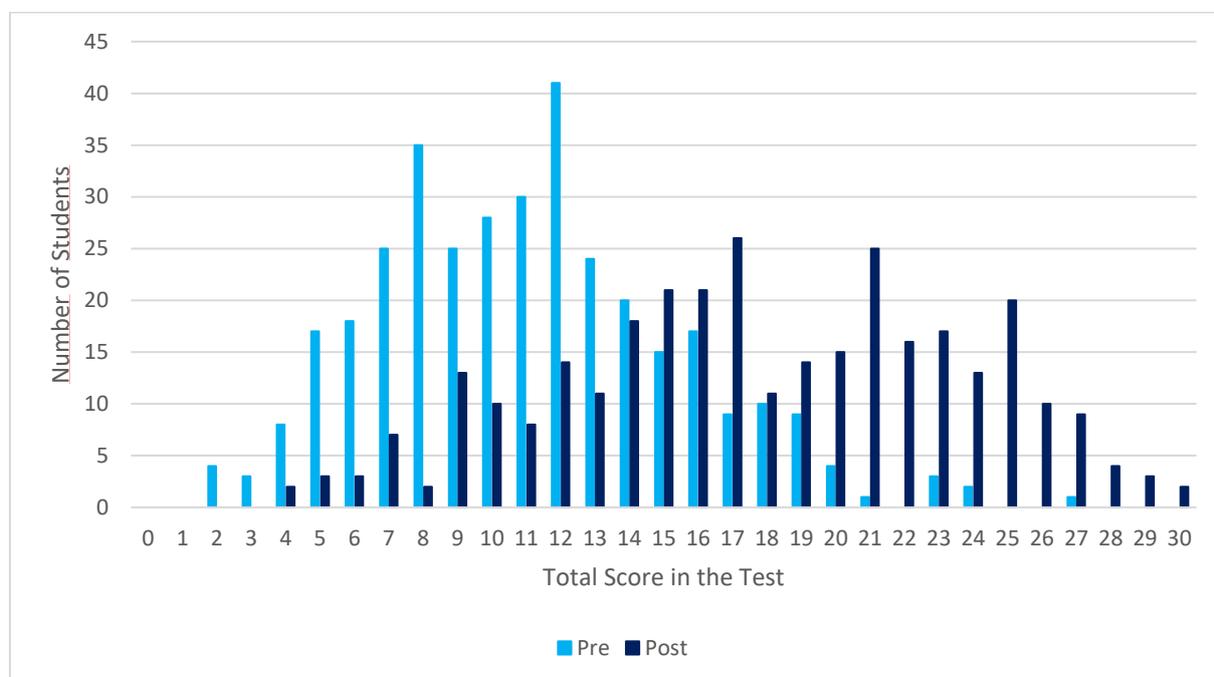

FIG. 9. Histogram of students' total test score (without item Q2) for the pre- (light blue) and posttest (dark blue).

Figure 9 shows that the test does not show floor effects of testing: No student scores 0 or 1 point in the total score, not even in the pretest; it shows ceiling effects of testing in the posttest for only 2 out of 318 students (0.63%) who achieved the maximum test score. Therefore, the test can differentiate well between different levels of students' conceptual understanding both before and after the teaching of converging lenses at school.

### G. Correlation of students' grades and their test scores

As, in general, students' grades give a good insight into the overall performance of a student in a certain discipline, we used students' physics grades as a measure of students' general physics performance and compared the students' CLCI test scores to their grades. For this analysis, we again used the test scores of the final 15-item version of the test, excluding item Q2.

We calculated Pearson's correlation coefficient for students' pretest performance with their physics grades yielding $r = -.24$, $p < .0001$, and for each posttest and follow-up test performance with their physics grades $r = -.42$, $p < .0001$ (the minus arises because in the German grading system lower grades indicate better performance). This indicates that the post- and follow-up test CLCI scores have a good correlation with students' general physics performance, which further strengthens the evidence for the CLCI being a valid test instrument for middle school students. The lower correlation between grades with the students' pretest performance compared with their posttest and follow-up test performance is plausible, given the students' lack of prior knowledge of the topic of the test.

### H. Test-retest reliability

To evaluate the test-retest reliability of the CLCI, we used the paired data of pretests and posttests as well as posttests and follow-up tests. We again used the test scores of the final 15-item version of the test, excluding item Q2. We calculated the Intraclass Correlation Coefficient (ICC) with a two-way mixed-effects model, single measurement type, and absolute agreement definition as proposed for test-retest reliability evaluation by Koo and Li [32]. The ICC for pretests and posttests was κ = 0.19 (95% CI = –0.03; 0.37); $F = 2.1$, $df1 = 276$, $df2 = 276$, $p < 0.0001$. This indicates a poor reliability (values less than 0.5), according to Koo and Li [32]. The ICC for posttests and follow-up tests was κ = 0.54 (95% CI = 0.38; 0.66); $F = 3.5$, $df1 = 71$, $df2 = 71$, $p < 0.0001$, indicating a moderate reliability (values between 0.5 and 0.75), according to Koo and Li [32]. It is appropriate that the ICC for pre- and posttest is poor and smaller than the ICC of the post- and follow-up test, because students' conceptual understanding improved significantly from pretest to posttest. Against the background of the rather long time period between the posttest and the follow-up test and the potential small change in conceptual ideas of the students (e.g., through forgetting or further physics instruction), the CLCI's test-retest reliability for post- and follow-up test can be considered sufficiently good.

## V. QUALITATIVE VALIDATION

For our qualitative analyses on construct validity, we used student interviews and expert surveys. Student interviews were conducted to gain an insight into students' actual thoughts about and understanding of the items and the answer choices, and why students decided for or against an answer choice. The expert survey was conducted to confirm the curricular validity and content validity of the test for the topic of converging lenses taught in middle school.

To strengthen and complement the findings from the quantitative analyses and to further evaluate the construct validity of the CLCI, we went into more detail using qualitative analyses. We chose two methods: conducting student interviews and conducting an expert survey.

### A. Student interviews

#### 1. Method of the student interviews

We recruited students for interviews from classes that had already been taught about converging lenses. Students were offered a 10€ reimbursement for participating in an interview; they had the opportunity to contact us via their teacher if interested. We accepted all seven students who were interested in participating in interviews. Demographic and other background data about the seven students is provided in Table VI.

TABLE VI. Demographic and other background data about the students participating in the interviews. Students' names were changed for reasons of data protection.

| No. | Name | Grade | Age | Gender | Mother tongue | Physics grade |
|---|---|---|---|---|---|---|
| 1 | Alexander | 8 | 13 | Male | German | 3.1 |
| 2 | Ben | 10 | 15 | Male | German | 2 |
| 3 | Charlotte | 7 | 13 | Female | German | 1 |
| 4 | Daniel | 7 | 13 | Male | German, Russian | 2.3 |
| 5 | Emma | 7 | 12 | Female | German | 1.5 |
| 6 | Fleur | 7 | 13 | Female | German | 1 |
| 7 | Gabriel | 7 | 13 | Male | German | 1.3 |

The interviews were scheduled by the students and were conducted online via one-on-one videocalls with only the leading researcher and the student participating. In the interviews, the students' screens with the test questions and the conversation audio were recorded. Students and their parents were required to give consent about the recordings and the students' voluntary participation in the interview. During the interviews, the researcher introduced herself and explained how the interview would proceed. Then, the students shared their screens and opened the online test via a link provided by the researcher. The students answered the demographic and background questions first without being interrupted by the researcher. Then, the researcher explained the format of the CLCI and asked the students to express what came to their minds when solving a question (think-aloud); regarding the test question, the single answer options were why they would choose a certain answer option and why they would not choose the others. When the student solved the test question by question and think-aloud, the researcher asked small encouraging questions, asking if the student had stopped to think aloud or asking for a more elaborate explanation if the student's explanation for why he/she did or did not choose an answer option was very short, superficial, or lacking.

## 2. *Results of the student interviews*

The student interviews provided very informative insights into how students understood the items and the answer options and how they decided for one or the other answer option. The complete interview transcripts can be found in the Supplemental Material I [33]. Due to space restrictions, only selected examples can be used in this section to illustrate the main findings from the student interviews.

All participants explained their answers to all 16 items except for participant Ben, who only solved 11 items out of the test for time reasons (his test lacked Q1, Q2, Q3, Q13, and Q15). The sample of participants consisted of a good mixture of high- and low-achieving students when considering their total scores in the test: while Alexander, Ben, and Daniel achieved rather low total scores (A: 28%, B: 18%, D: 31%), Charlotte, Emma, and Fleur scored rather high (C: 88%, E: 72%, F: 67%). Gabriel scored mediocre with 44%. This distribution is in line with the students' self-reported physics grades (cf. Tab. 3); the Spearman correlation coefficient of students' total test scores and their physics grades was $\rho = -.76$, $p = .049$. The item that was solved correctly by most students was Q7 (6 out of 7); the item that was solved incorrectly by most students was Q15 (0 out of 6). Q14 was also frequently solved correctly (5 out of 7); Q13 was frequently solved incorrectly (1 out of 6). These results are in line with the item statistics according to the psychometric analyses reported above and the experts' survey answers described below.

The students' reasoning for selecting incorrect answer options matched well with the student difficulties that were used for creating the distractors during the instrument development. Likewise, students' reasonings for selecting correct answer options matched well with the correct conceptual ideas that were intended to be represented by these answer options. Some examples for reasoning regarding incorrect and correct answer options are provided in Table VII.

TABLE VII. Examples for students' reasoning about why they chose a specific answer option. 'CC / SD' stands for "core concept" or "student difficulty" addressed in the student's answer.

| Item | CC / SD | Student reasoning |
|---|---|---|
| Q3 | HV | Charlotte (correct): "I would say that the observer can see the duck in D and also in C. I wouldn't say the same for A and B, because in B the light does not reach the duck at all; in A it doesn't either... (*unintelligible murmuring*) ...so definitely C and D." |
| Q5 | TB1, TB | Alexander (incorrect): "It's B. Because I have watched an explanatory video and there was also an example like that. For example, the boy is now upright and on the other side he is … um … upside down … ah … what about D? No … um … it's D, because now that I've looked again, it's D because with B it would be left-right-reversed and that wasn't the case, so now I'm logging in D."<br><br>Ben (incorrect): "It would then be flipped … um … like in a mirror, so I don't know why it would be upside down at all. [...] I would simply exclude these [means B and D] because otherwise it would have to be ... turned around ... again ... somehow. The rays are actually just sent to the other side. That's why I think it's C." |
| Q8 | CS | Fleur (correct): "Well, I think that if you were to start from this, for example, from the arrowhead, um, then the rays would first, the rays would first be refracted on the focal plane and then go downwards and I think that because the screen is farther away, they have a longer path and therefore they also go lower at some point, because, for example, if they were to go straight and then be refracted, then they wouldn't stop at some point and then kind of go on horizontally, but I think that they simply continue diagonally and, so to speak, the farther back the screen is, the larger the object, well the image, then." |
| Q9 | SR | Emma (correct): "I would say that it's B, because um that ... um ... I think you just can't see the image anymore if the screen doesn't catch the image, but, so, I think that the screen makes it visible, except that it's still there, just not in the same place, so that you just can't see it anymore." |
| Q10 | DL | Daniel (incorrect): "B would be right for me, because the lens would be larger and the object is also in that size and then the image on the screen is larger as well, because with a smaller lens, the image would also be smaller here. And because it's a bigger lens now, I think the image point becomes bigger to the arrow."<br><br>Gabriel (incorrect): "Um, so the rays kind of hit ... so the rays are sort of emitted and they arrive at the lenses at different points. And the topmost and the very lowest are refracted accordingly. And the larger the lens, the further up and the further down the ray is then refracted and therefore does not arrive at the same place as before, but further down and further up. And that's why I would say it gets bigger [chooses B]." |
| Q11 | DL | Daniel (incorrect): "That is because of the lens. If it were larger, then the image on the screen would also be larger, and when it's smaller, then the image on the screen is also smaller [chooses C]." |
| Q12 | PC1 | Ben (incorrect): "D doesn't make any sense because, for example, if the tip passes through the focal point, it will bounce off the aperture down here and then be reflected here and wouldn't even reach the lens. E seems reasonable, because for example, here, where the arrow kinda stops, it still passes through, no problem, and can also still be reflected and then arrive at the back. […] That [pointing to C] would also make no sense, because then it is more or less the same as if there were no lens at all, that this would simply be an image |

| Item | CC / SD | Student reasoning |
|---|---|---|
| | | of an arrow behind it, or that there is the lens in the middle, then the whole thing is refracted again and then taken apart again, so to speak. And that's why I would say E, because for example here the tip doesn't fit through the pinhole and is then deflected. So, example here below, that a part passes through the aperture again and then arrives down there." <br><br> **Daniel (incorrect):** "I think that A and D, so that A is wrong, that you don't see anything, because you could see something for sure. And that D is also wrong, because it can't actually remain unchanged. Something has to change there. I also don't think that the image becomes fainter. I simply think that the image … you only see the middle part and cannot see the arrow, that the edges are cut off." <br> **Interviewer:** "Then the question is, cut off as in C or as in E?" <br> **Daniel:** "I think it would be ... Fifty, fifty, E." <br> **Interviewer:** "What would be your reasoning?" <br> **Daniel:** "Mmh. Because it could also be enlarged by the lens, the middle part." |

A structural analysis of the students' answer behavior suggests that the structure which was found and confirmed with the structural analysis (EFA and CFA) appears here, again. There are noticeable parallels within the answering behavior of the students, for example, for items Q12 and Q14, where 6 out of 7 students chose either the correct answer option in both items or chose the distractors corresponding to the same underlying student difficulties in both items. The same holds true for items Q10 and Q11, where 5 out of 7 students chose the respective corresponding answer options, and Q5 and Q6 for 4 out of 7 students. Items Q4 and Q9 were both answered correctly or incorrectly consistently by 6 out of 7 students, and the item cluster of Q8, Q15, Q16, and Q13 was answered correctly or incorrectly almost consistently by 3 out of 6 students (when neglecting one of the three correct answers in the very difficult item Q15).

When looking into the students' answers for item Q2 in more detail, it appears again that this item might be misleading and problematic in how the students interpret the answer options. They choose the incorrect answer option A because they "have also drawn something like this before" (Daniel) and this is "the way [they] always drew light rays in physics class […] around the light source" (Charlotte). They say that "the light bulb is the center point, so to speak, and then it spreads out" (Emma) and "light is a center point, it spreads out evenly in all directions" (Gabriel). Gabriel excludes the correct answer C because "the light comes ... the light bulb containing it, this doesn't bring the light to the outside, but it is the power within the light bulb." These answers suggest that the students do not see the light bulb as an extended light source, but as a point light source. They are familiar with the concept of light bulbs being point light sources and therefore did not interpret the answer options of this question correctly. This, again, strengthened the evidence for removing item Q2 from the CLCI.

### B. Expert survey

#### 1. *Method of the expert survey*

The expert survey consisted of a Microsoft Word document in which the CLCI, all questions about the CLCI, and the corresponding answer fields were embedded. We asked five experts from the field of physics education in research and practice (research associates who work as teachers at the same time) to participate in the expert survey. They had to solve the test and give reasons for each of their choices of answer options. Then, they were asked to evaluate the single items of the test and judge them according to the criteria of: (1) how relevant they were to the middle school curriculum (curricular/content validity) on a scale of 1 = *not relevant at all* to 5 = *very relevant*, (2) how suitable they are for the students' age group/grade (target group validity) on a scale of 1 = *not suited at all* to 5 = *very suited*, and (3) for which learner level in the target group they would suggest the item (i.e., how difficult they rate the item) on a scale of 1 = *learner level below average*, 2 = *learner level is average*, and 3 = *learner level above average*. The experts were also asked to provide their brief reasoning for their judgement of each item. After rating all 16 items individually, the experts were asked to judge the overall suitability of the test, again on the scale of 1 = *not suited at all* to 5 = *very suited* and they were again asked for their brief reasoning for their judgement.

#### 2. *Results of the expert survey*

The experts' ratings for the single items and the test in total for the three categories of interest are shown in Table VIII. The experts' written reasoning and evaluations about the items can be found in Supplemental Material II [33].

TABLE VIII. Means and standard deviations of the experts' ratings for the single items and for the test in total. Item difficulty was taken from the quantitative analyses and is displayed for easier comparison with the Learner's level.

| Item | Core concept | Relevance [1:5][a] | Suitability [1:5][a] | Learner's level [1:3][b] | Item difficulty [0:1][c] |
|---|---|---|---|---|---|
| Q1 | LP | 4.4 (0.5) | 4.0 (0.6) | 1.9 (0.7) | .623 |
| Q2[d] | LS | 4.5 (0.9) | 3.25 (1.5) | 2.0 (0.7) | .431 |
| Q3[d] | HV | 5.0 (0.0) | 4.25 (0.8) | 2.1 (0.2) | .708 |
| Q4 | RP | 5.0 (0.0) | 4.8 (0.4) | 1.9 (0.5) | .668 |
| Q5[d] | TB | 4.8 (0.4) | 4.5 (0.5) | 2.1 (0.7) | .624 |
| Q6 | TB | 4.4 (0.8) | 3.8 (0.7) | 2.3 (0.7) | .730 |
| Q7 | PI | 4.8 (0.4) | 4.2 (0.7) | 2.1 (0.7) | .868 |
| Q8 | CS | 4.8 (0.4) | 4.6 (0.5) | 2.5 (0.4) | .571 |
| Q9 | SR | 4.4 (0.5) | 4.6 (0.5) | 2.3 (0.7) | .665 |
| Q10 | DL | 4.4 (1.2) | 4.0 (0.9) | 2.1 (0.7) | .616 |
| Q11 | DL | 4.4 (1.2) | 4.0 (0.9) | 2.1 (0.7) | .615 |
| Q12 | PC | 3.6 (1.4) | 4.2 (0.7) | 2.6 (0.5) | .454 |
| Q13[d] | OL | 4.5 (0.9) | 4.3 (0.8) | 2.4 (0.4) | .329 |
| Q14 | PC | 4.2 (1.2) | 4.4 (0.8) | 2.6 (0.5) | .777 |
| Q15 | RM | 4.0 (1.5) | 4.2 (0.7) | 2.8 (0.4) | .283 |
| Q16 | VI | 4.4 (0.8) | 4.4 (0.8) | 2.0 (0.9) | .390 |
| Test in total | All concepts | n/a | 4.2 (0.4) | n/a | n/a |

[a] Relevance and Suitability were indicated on a scale of 1 = *strong disagreement*, 2 = *disagreement*, 3 = *undecided*, 4 = *agreement*, and 5 = *strong agreement*.
[b] Learner's level was indicated on a scale of 1 = *below average*, 2 = *average*, and 3 = *above average*.
[c] Item difficulty was indicated by 0 = *low solving probability*, 1 = *high solving probability*.
[d] These items were each answered incorrectly by one of the experts, whose ratings were therefore excluded from the summary.

The experts' reasoning for choosing the correct answer confirmed the reasoning that was intended when the items were created. All items show a good rating for relevance (4.0 or higher), except for item Q12, which was rated with 3.6 on average. For this item, the experts either strongly supported Q12's relevance (two experts rated it with 5, one with 4) or judged it as not relevant (two experts rated it with 2). The reasons for rating the item with 2 were: "The question presupposes a deeper understanding of the visual process and the formation of the image (since, for example, it is no longer possible to draw in the focal ray)" and "I consider the combination of pinhole and converging lens to be of little relevance." The reasons for rating the item with 4 or 5 were, for example, "Difficult question, but very relevant. Requires deeper understanding of the topic" and "This question is very relevant to evaluate the understanding of how an object is imaged through a lens." According to all experts, this item requires deeper understanding. Therefore, the critical question is whether this deeper understanding is relevant; we answered with "yes" for this test and therefore kept item Q12 in the test.

All items were rated well-suited (4.0 or higher), except for items Q2 (rated 3.25 on average) and Q6 (rated 3.8 on average). Item Q2 was rated as suited (4 and 5) by two experts; one expert rated it with 3, and one expert with 1. The expert rating the item as "undecided" in suitability reasoned: "I find the item difficult to assess. From textbooks, etc., one would always assume a point light source and therefore prefer A as the answer. It therefore also depends somewhat on the light source assumed." The expert rating the item with 1 reasoned: "A and C problematic. To solve this item correctly, an extended, diffuse light source MUST be taught. Otherwise, it is not possible to distinguish whether the lamp is only symbolic, like in the majority of sketches in class (in which case, answer A would clearly be correct). Also, it is not clear if the inside of the lamp absorbs, as in the case of the sun, in which case A could also be considered correct. Third, the projection from 3D to 2D could lead to confusion. Do the rays in C go to the inside of the lamp or pass it?" The expert rating with 4 still commented: "I find the illustration difficult because it neglects the internal structure of the bulb." Only the expert who rated the item with 5 wrote: "If an object is imaged by a converging lens onto a screen, the object can be perceived as an extended light source. This is important for the basic understanding of the construction of ray paths." The experts' concerns about item Q2 were also echoed frequently by the teachers of the classes participating in the pre-post assessment for the psychometric analyses of the test and also occurred in the student interviews. This, again, supports the argument to remove item Q2 from the test.

Item Q6 was criticized by the experts mostly because of its similarity to item Q5; however, it was described to be more abstract than Q5 by the experts. Therefore, to keep the differentiation of abstraction within the conceptual idea represented in items Q5 and Q6, we decided to keep item Q6 in the test.

The questions' difficulties, indicated by the rating of Learner's level, seem to be appropriate according to the experts. The difficulty rating showed values between 1.9 and 2.8, with nine items between 1.9 and 2.1, indicating average Learner's levels. The items related to higher learner's levels (item Q15, Q12, Q8, and Q13) also showed higher difficulty in the psychometric analyses of the test, with item Q15 appropriately evaluated as the most difficult item. Only items Q2 and Q16 were rated as easier by the experts compared to the difficulty shown in the psychometric analyses, and item Q14 was rated as more difficult than it turned out in the psychometric analyses.

## VI. DISCUSSION, LIMITATIONS, AND OUTLOOK

### A. Discussion

In this paper, we describe the development and validation process of the Converging Lenses Concept Inventory (CLCI). We collected evidence for the degree to which this test instrument can comply with the three claims for concept inventories presented in the introduction. These claims were that a CI can assess: (1) the overall understanding of all concepts within the test, (2) the understanding of specific concepts, and (3) students' propensity for difficulties within this topic [9].

There are eight facts that support Claim 1 (measuring the overall understanding of the concepts):

First, the distractor analyses show that the distractors were chosen by the students much less often in the posttest than in the pretest. Second, the analyses according to classical test theory after removing item Q2 show "good" values for the item statistics difficulty and discrimination, an "average" value for Cronbach's alpha of the total score, and "excellent" values for Cronbach's alpha-with-item-deleted, according to the categorical judgement scheme by Jorion and colleagues [9]. However, the fact that Cronbach's alpha of the total score was average is not surprising for a test instrument that consists of multiple subscales [9]. Third, analyses according to item response theory revealed an "excellent" for the individual item measures of the test, because all items fitted the model [9]. Fourth, the test scores are normally distributed in both pre- and posttest and a clear shift from the pretest to the posttest total scores distribution can be observed. Fifth, the students' post- and follow-up test scores show a good correlation to their general physics performance (measured in physics grades). Students' physics grades are very likely an indicator of their overall conceptual understanding in physics, thereby also attesting to the criterion validity of the test. Sixth, the test-retest reliability of post- and follow-up test was moderate and higher than the test-retest reliability of pre- and posttest. Concepts can change due to instruction (pre-post), but if not further addressed, they can be considered rather stable constructs (post-follow-up). Still, it is not known what happened in between the two test points of posttest and follow-up test. Perhaps more teaching about converging lenses or optics took place and slightly modified the students' conceptual understanding of the topic assessed in the CLCI, or maybe students simply forgot about the topic over time. This can explain why the test-retest reliability of the post- and follow-up test is only medium. Seventh, the student interviews suggest that the test is able to assess the overall understanding of all concepts within this topic, suggested by the good correlation of students' performances in the test and their physics grades. Eighth, the experts' ratings suggest that the single items are highly relevant for this topic and the single items as well as the overall test are well-suited for assessing middle school students' conceptual understanding for image formation by converging lenses.

Moreover, there are four facts that support Claim 2 (adequately measuring the understanding of specific concepts):

First, concerning the structural analyses, we found in the exploratory factor analysis that the exploratory factor constructs also conformed very well to the subscales we used for the test development. This can be rated "excellent," according to Jorion and colleagues [9]. Second, the confirmatory factor analysis showed "good" values for item loading and "excellent" values for the comparative fit index and the root-mean-square error approximation. Third, the subscale Cronbach's alpha values can be considered good according to the circumstances of small subscales. Therefore, these subscales show good reliability in measuring the conceptual understanding of the respective core concepts. Fourth, the student interviews, especially the findings about the structures in students' answer choices and argumentations that fit very well to the previously detected factor structures, suggest that the test can measure the understanding of single concepts, as well.

Furthermore, there are three facts that support Claim 3 (detecting students' propensity for difficulties):

First, the distractor analysis showed that all distractors seem to be fair alternatives to the correct answer options, especially in the pretest, indicating that the students' difficulties we intentionally built into the distractors work well and are a good indicator of whether this specific difficulty is possessed by the respective student. The fact that students tended to select the correct answer options after they were taught about converging lenses supports, again, that the CLCI can measure students' difficulties, because these difficulties should be decreased after instruction about the topic. Second, the good fit of the underlying core concepts with related student difficulties and the structural model detected in the structural analyses suggests that this test instrument can evaluate students' difficulties well. Third, in the student interviews, students' reasonings about why they chose a distractor based on the underlying students' difficulty provided detailed insights into their actual conceptual understanding and their erroneous conceptions. Here, students' answers indicate that the test can detect students' propensity for difficulties.

All in all, this combined evidence attests to a good validity of the CLCI, if item Q2 is removed from the test. As this item is not immediately relevant content-wise, as it does not include a conceptual idea about a converging lens, but rather about light propagation, this basic item can be removed without the need to replace it.

### B. Limitations

As the CLCI was developed as a parsimonious one-tier instrument to be used in schools and research settings, we cannot directly assess students' response confidence and their explanation for selecting the respective answer option. If such information is important to the teacher or evaluator, the additional tiers can, of course, be added to the single test items.

Moreover, the CLCI is a multiple-choice test, which, among the advantages of this format, also highlights the shortcoming that students cannot indicate their actual thoughts in open answers; they can only choose from the given answer options. Students' underlying reasonings for answers and their underlying conceptions can, therefore, not been assessed in the same way as with open answer tests or in interview situations.

Additionally, the differences in test administration should be mentioned at this point; pre- and posttests were administered in-person, but the follow-up tests were administered via an online setting. It is not clear whether these different test settings might have had an influence on students when they solved the test. However, we do not expect differences, given the good test-retest reliability of posttest and follow-up test, the similarities of answer choices in post- and follow-up tests, and the similar duration of the tests when students took them.

An additional weakness of the test is based on the partial credit model we used to score students' answers: If students chose all answer options in all items, they would theoretically get partial credit for them all. However, this was not the case for a single student participating in one of our tests. Even selecting all answer options for a single item occurred rarely in our sample. If this occurs, test administrators can still decide whether this might be a systematic method to collect partial credit points in the test and score the items answered in this manner with 0 points, or whether the student might actually have perceived all answer options as correct and, therefore, correctly gets the partial credit.

Moreover, the partial credit model used for scoring students' tests makes items with more than one correct answer option more difficult than items with only one correct answer option. This could be one reason why item Q15 was so difficult for students. However, item Q3 showed a medium difficulty. One solution to these shortcomings of the partial credit model could be to transform the CLCI into a dichotomous test instrument by cutting one of item Q3's correct answer options and making multiple items out of the initial item Q15 with only one correct answer within each. This might be subject to further analyses of the test instrument. However, we want to emphasize at this point that we intentionally announced to students that each item could have multiple answer options to encourage them not to rush over the test, but to properly read all presented answer options and to consider whether multiple suggested answer options could be correct in their perception.

### C. Outlook

Overall, the 15-item version of the CLCI can be recommended as a valid and comprehensive test instrument to assess middle school students' understanding of image formation by a converging lens. It can be used as a formative or summative assessment in school or in research settings. Concept inventories are always subject to development, further improvement, and change [9], as is the CLCI; further implementation of the CLCI, further discussion with experts, and more data of students' answers can help, in the future, to eventually improve the items and their answer options even more. For instance, the CLCI could be transformed into a dichotomous test instrument, as suggested in the limitations, and the potential of this transformation for improving the CLCI as a test instrument could be evaluated. Moreover, as the items are constructed with multiple external representations (i.e., texts and pictures), more process-related analyses of the items would be interesting, for example, investigating the information selection and information integration processes of this item format via eye-tracking. In this context, the interplay of students' representational competence [34] and their understanding of the test items might also give new insights into how the CLCI can best be used to assess students' conceptual understanding of image formation by converging lenses.

**ACKNOWLEDGEMENTS**

We want to thank Sebastian Zangerle, who helped substantially with the graphical design of the pictures included in the items and with the item development.

Salome Wörner is a doctoral student at the LEAD Graduate School & Research Network, which is funded by the Ministry of Science, Research and the Arts of the state of Baden-Württemberg within the framework of the sustainability funding for the projects of the Excellence Initiative II. In addition, she is a junior fellow within the program "Kolleg Didaktik:digital" at the Joachim Herz Stiftung (Hamburg, Germany).

# APPENDIX A: ORIGINAL FULL TEST – 16 ITEMS

**Test instructions for students prior to solving the CLCI:**

"You need around 20 minutes for this test. It is important that you can work undisturbed and concentrated during this time.

Most questions include answers with text and pictures. Always look closely at the picture and read the text carefully. Answer the questions to the best of your knowledge. If you are unsure about a question, you can also decide intuitively which answer or answers are correct.

It is important for us that you answer the questions as well as you can (but without aids such as books or the Internet!) and that you don't simply always choose 'A.' Read all answer options properly!

The test's result is NOT counted into your physics grade.

Don't panic, it is okay if you check something wrong or if you don't know an answer!"

# Converging Lenses Concept Inventory

This test only contains multiple-choice questions.

In several questions **more than one answer option can be correct**. Select **all correct answers** in this case.

You receive **2 points** for correctly solving the item, **1 point** for a partly correct solution and **0 points** for an incorrect answer.

**Example:**

Which of the depicted animals is a mammal?

*You receive 2 points for selecting answers A and D.*

*You receive 1 point for, e.g., selecting answers A and B.*

*You receive 0 points for selecting neither answer A nor answer D.*

In case you ticked a box but change your mind later on, please cross out your previous answer completely to make your new choice obvious.

**Example:**

**Stop!
Don't turn the page yet.
Please wait until we
continue together.**

# CONVERGING LENSES CONCEPT INVENTORY

**Question 1**

Which properties of light are correct?

A
☐ Light is a kind of substance (similar to a fluid) that immediately fills rooms when the light source is turned on.

B
☐ A light source consists of quite a lot of light that is held in one place.

C
☐ Light, if not shielded or deflected, spreads farther and farther from the light source.

D
☐ When two light beams overlap, the light rays from the light beams collide and can even bounce off each other.

**Question 2**

How do light rays propagate from the surface of an extended light source (e.g., an incandescent lamp)?

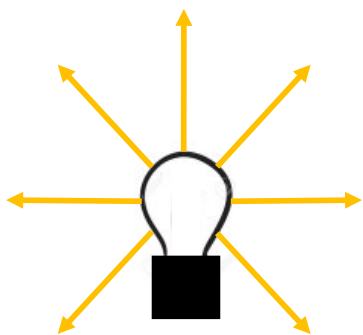 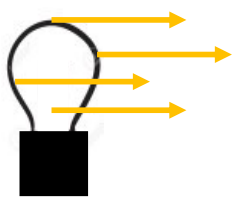 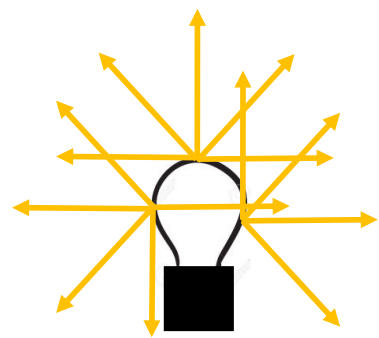

A
☐ The light rays only propagate outward from an imaginary center of the extended light source.

B
☐ The light rays propagate from different points on the surface only in the direction in which the light should go (e.g., to the right).

C
☐ The light rays propagate from any point on the surface in any direction.

**Question 3**

An observer is in an otherwise dark room with a normal, yellow (non-luminous) rubber duck, a light source and an opaque shield. The pictures show the situation from above. In which arrangement can the observer see the duck?

A ☐
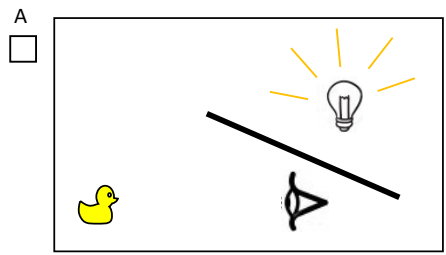

B ☐
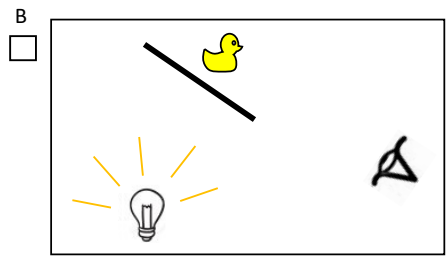

C ☐
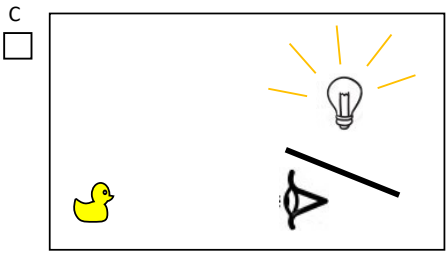

D ☐
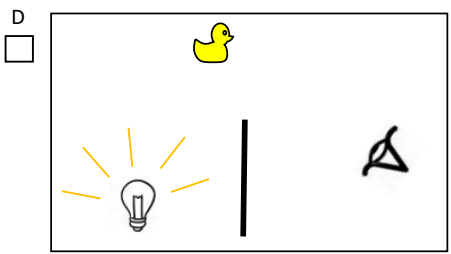

**Question 4**

Consider the illustrations that show the refraction of light through a converging lens. What happens to the light behind the lens?

A ☐
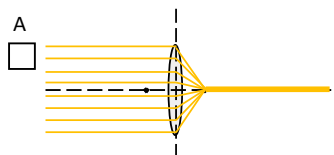

Behind the lens, the light rays converge in the focal point and continue to radiate behind it like a laser.

B ☐
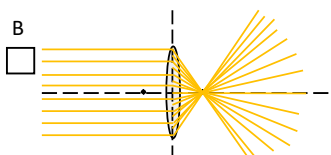

The light is multiplied by the lens. Behind the lens, all light rays pass through the focal point and then diverge again.

C ☐
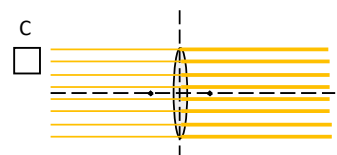

Behind the lens, the light rays are brighter. They are not bundled but continue parallel as they came.

D ☐
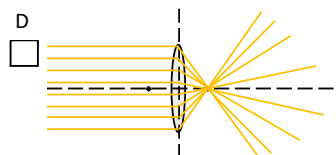

Behind the lens, the light rays pass through the focal point and then diverge again.

E ☐
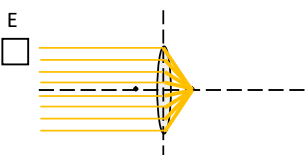

Behind the lens, the light rays are brighter. They are bundled in the focal point and do not radiate farther behind it.

F ☐
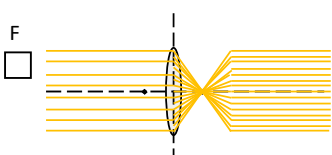

The light is multiplied by the lens. Behind the lens, all light rays pass through the focal point and then return to their parallel paths.

**Question 5**

An original image is generated in a projector and projected onto the screen on the wall through a converging lens at the front of the device.

Which of the images is the original image that was projected onto the screen using the lens ?

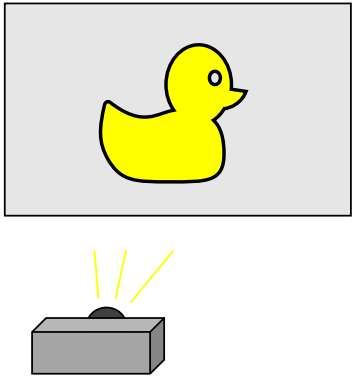

A
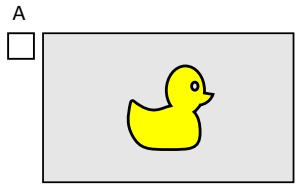
The duck is the same in the original image as in the image on the screen.

B
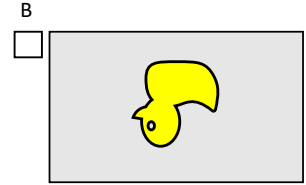
In the original, the left and right sides of the duck are reversed, and it is upside down.

C
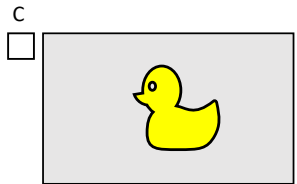
In the original, the duck is mirrored horizontally.

D
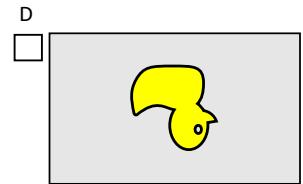
In the original, the duck is upside down.

**Question 6**

An original image is generated in a projector and projected onto the screen on the wall through a converging lens at the front of the device.

At which of the four positions on the circle must the dot in the original image be to create the correct image on the screen?

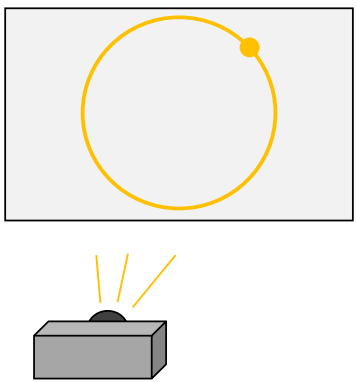

Image on the screen
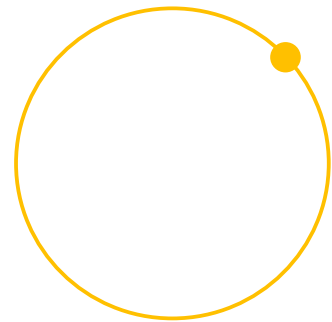

Original image in the projector
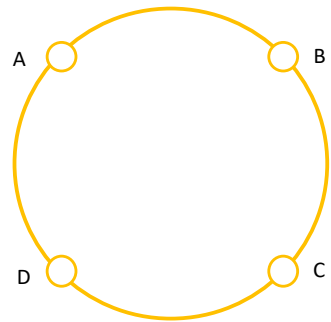

**Question 7**

A luminous object is to be projected in focus onto a screen using a converging lens. Which answers correctly show how the image is created?

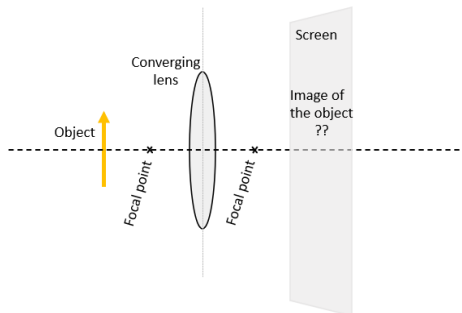

A 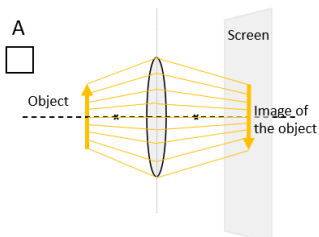

The light rays from the arrow travel side by side through the lens, where they are refracted and, as a result, create an inverted image.

B 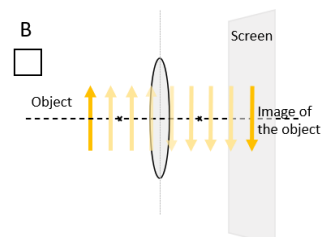

The arrow travels as a whole to the lens, is flipped by the lens, and then travels to the screen.

C 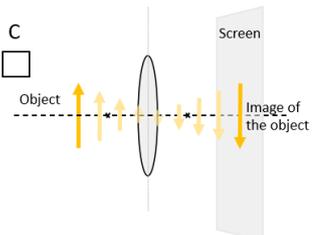

The arrow travels to the lens and becomes smaller; at the center of the lens it is flipped and then travels to the screen.

D 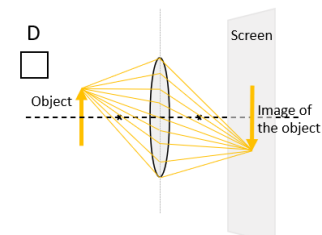

From every single point of the arrow (shown in the picture at only one point for better clarity), light rays travel through the lens, are refracted there, and, finally, travel to the screen.

**Question 8**

A luminous object is projected in focus onto a screen using a converging lens. Then the screen is moved away from the lens.
What happens to the image on the screen?

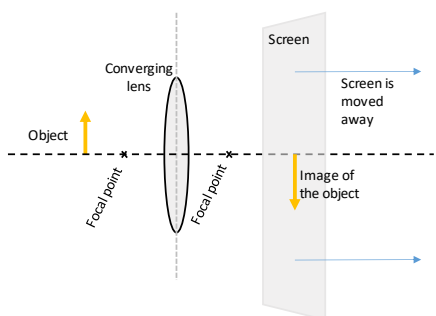

A 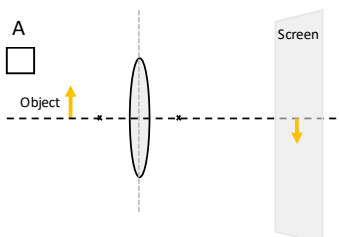

The image on the screen becomes smaller.

B 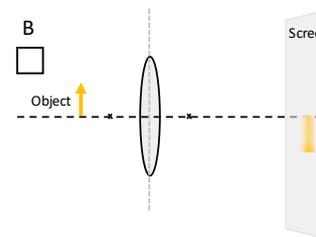

The image on the screen remains the same size and becomes blurred.

C 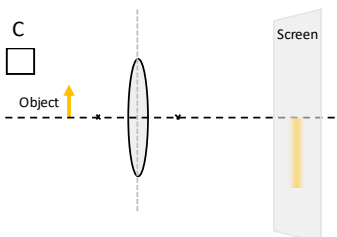

The image on the screen becomes larger and blurred.

D 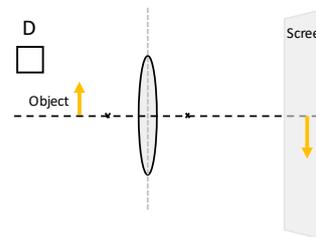

The image on the screen remains the same except that now the screen is farther away.

CONVERGING LENSES CONCEPT INVENTORY

**Question 9**

A luminous object is projected in focus onto a screen using a converging lens. What happens when the screen is removed?

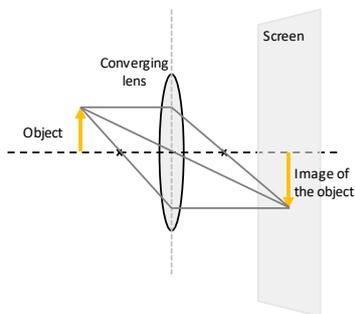

A
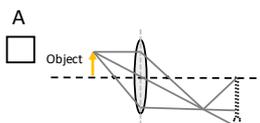
Without a screen, the image moves farther away because it was not caught earlier.

B
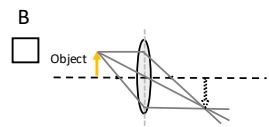
Without the screen, you just can't see the image anymore. It is still in the same place as before.

C
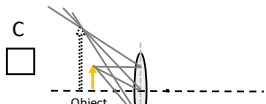
Without a screen, the rays are reflected by the lens.

D
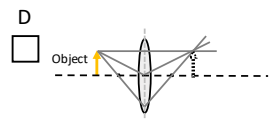
Without a screen, the lens no longer rotates the image.

E
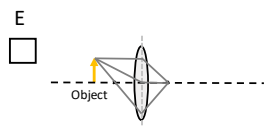
Without a screen, the lens only retains its collecting function. All rays end in the focal point.

F
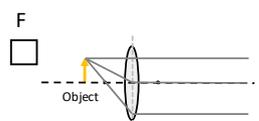
Without a screen, the lens does not "know" where to project the image. The result is an image at infinity.

**Question 10**

A luminous object is projected in focus onto a screen using a converging lens. How does the image change when a <u>larger</u> lens with the same focal length is used?

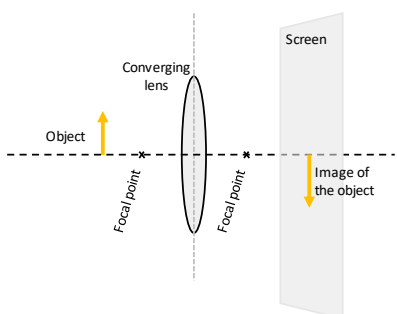

A
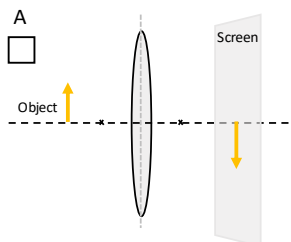
The image on the screen remains the same size.

B
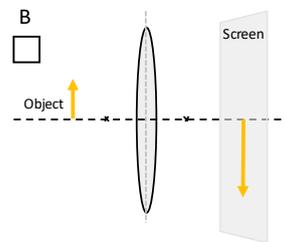
The image on the screen becomes larger.

C
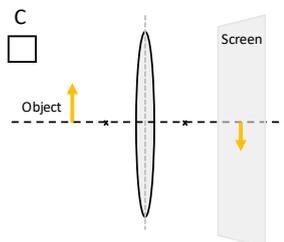
The image on the screen becomes smaller.

# CONVERGING LENSES CONCEPT INVENTORY

**Question 11**

A luminous object is projected in focus onto a screen using a converging lens. How does the image change when a <u>smaller</u> lens with the same focal length is used?

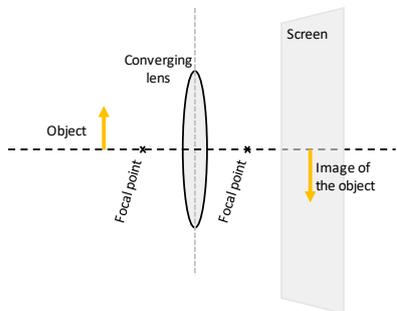

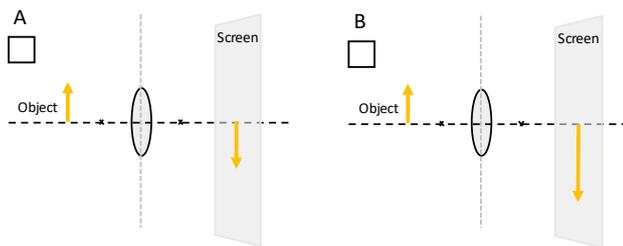

A — The image on the screen remains the same size.

B — The image on the screen becomes larger.

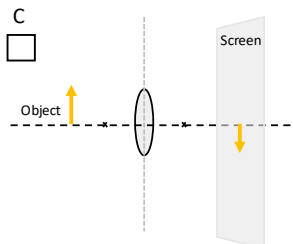

C — The image on the screen becomes smaller.

**Question 12**

A luminous object is projected in focus onto a screen using a converging lens. Then a pinhole is placed between the object and the lens. What happens to the image of the object?

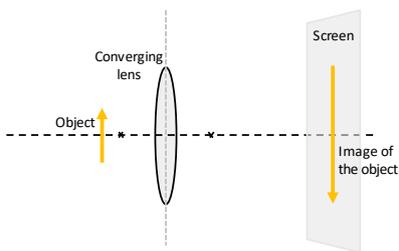

Pinhole from front    Pinhole from the side

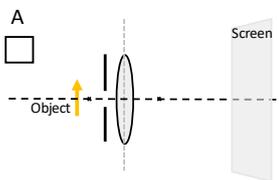

A — The image disappears because the arrow does not fit through the pinhole.

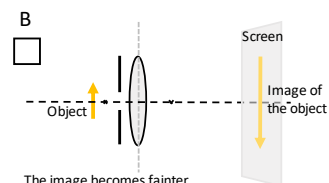

B — The image becomes fainter.

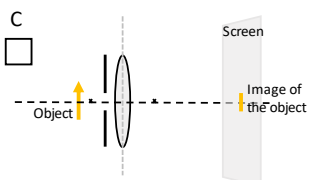

C — The image is cut to the size of the pinhole, you can only see the middle part.

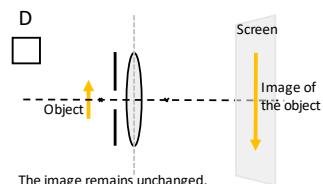

D — The image remains unchanged.

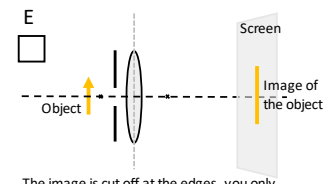

E — The image is cut off at the edges, you only see the middle part.

# CONVERGING LENSES CONCEPT INVENTORY

**Question 13**

A luminous object is to be refracted through a converging lens that is smaller than the object. What does the image look like on the screen?

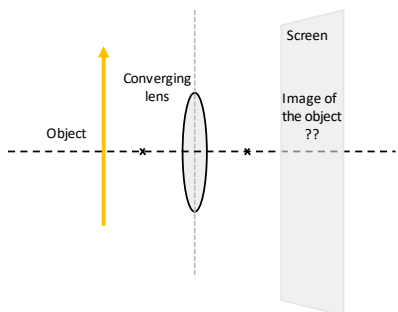

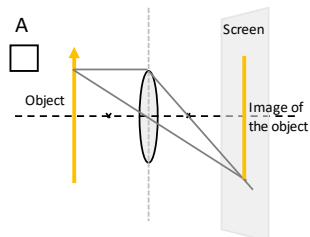

A

The tip of the arrow cannot be projected.

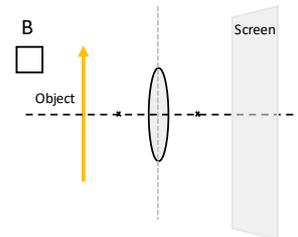

B

The object is too large for the lens, no image can be seen.

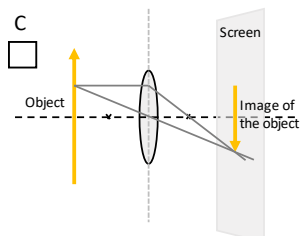

C

A smaller image of the arrow is created.

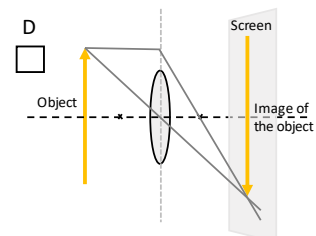

D

A larger image of the arrow is created.

**Question 14**

A luminous object is projected in focus onto a screen using a converging lens. What happens to the image of the object when a part of the lens is covered?

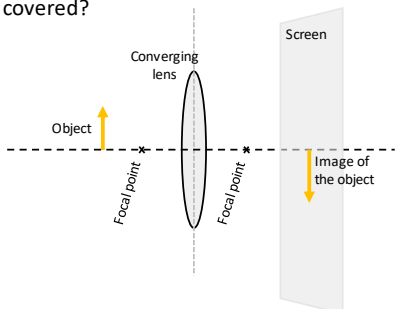

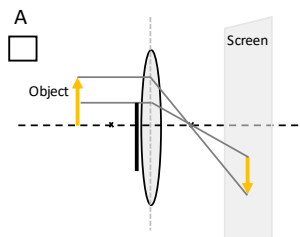

A

The lower part of the arrow cannot be projected.

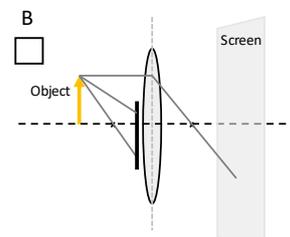

B

Only one construction beam passes the cover, therefore no image can be seen.

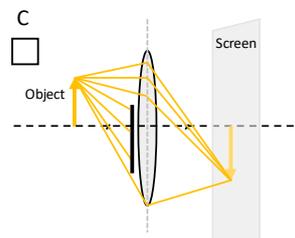

C

Fewer rays of light reach the screen, so the image is only fainter.

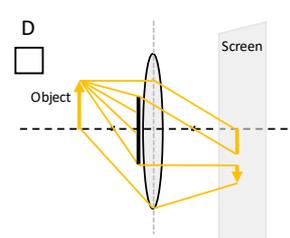

D

The middle part of the arrow cannot be projected.

**Question 15**

A luminous object is projected in focus onto a screen using a converging lens. Then the object is moved. Where do you have to put the screen in each case to get a sharp image and how big does the screen have to be to fit the image on it?

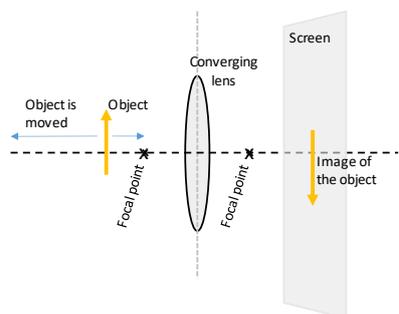

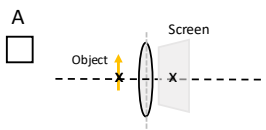
A

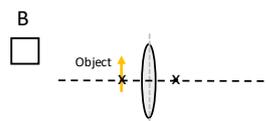
B

If the object is in the left focal point, then the image will be in the right focal point. You only need a screen that is slightly larger than the object.

If the object is in the left focal point, then the image will be at infinity. You cannot see it on a screen.

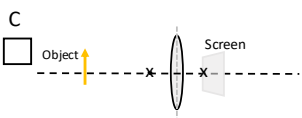
C

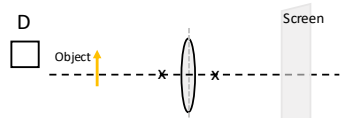
D

If the object is a bit farther away, the image will be closer to the lens. The screen can be smaller than the object.

If the object is a bit farther away, the image will be farther away from the lens. The screen needs to be large because the image will be large.

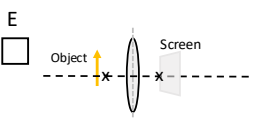
E

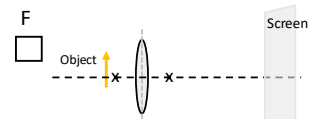
F

If the object is a little closer to the left focal point, the image will be closer to the lens. The screen can be smaller than the object.

If the object is a little closer to the left focal point, the image will be farther away from the lens. The screen needs to be large because the image will be large.

**Question 16**

If the object is closer to the converging lens than the focal point, a virtual image is created. Where is the virtual image constructed correctly?

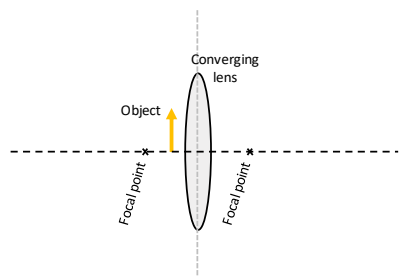

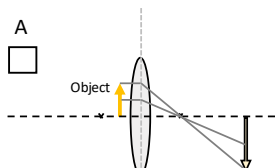
A

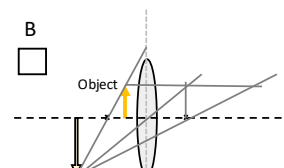
B

The virtual image is on the other side of the lens and is upside down.

The virtual image is on the same side of lens as the object and is upside down.

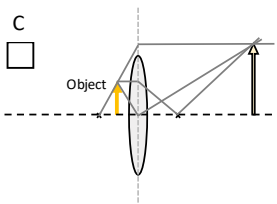
C

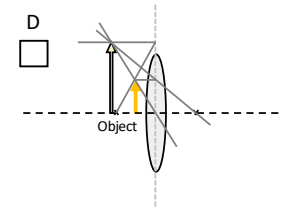
D

The virtual image is on the other side of the lens and is upright.

The virtual image is on the same side of lens as the object and is upright.

## APPENDIX B: EXTENDED INFORMATION ON TEST DEVELOPMENT

TABLE B.I. Format of the test items. "Default design" means that both the item stem and all answer options contained text and a picture.

| Item | Specific item design? | Number of answer options | Correct answer(s) |
|---|---|---|---|
| Q1 | Item stem w/o picture and answers w/o pictures | 4: A, B, C, D | C |
| Q2 | Item stem w/o picture | 3: A, B, C | C |
| Q3 | Item stem w/o picture and answers w/o text | 4: A, B, C, D | C, D |
| Q4 | Item stem w/o picture | 6: A, B, C, D, E, F | D |
| Q5 | Default design | 4: A, B, C, D | B |
| Q6 | Answers w/o text | 4: A, B, C, D | D |
| Q7 | Default design | 4: A, B, C, D | D |
| Q8 | Default design | 4: A, B, C, D | C |
| Q9 | Default design | 6: A, B, C, D, E, F | B |
| Q10 | Default design | 3: A, B, C | A |
| Q11 | Default design | 3: A, B, C | A |
| Q12 | Default design | 5: A, B, C, D, E | B |
| Q13 | Default design | 4: A, B, C, D | D |
| Q14 | Default design | 4: A, B, C, D | C |
| Q15 | Default design | 6: A, B, C, D, E, F | B, C, F |
| Q16 | Default design | 4: A, B, C, D | D |

# APPENDIX C: EXTENDED QUANTITATIVE DATA

**Distractor analysis:**

TABLE C.I. Percentage of students who chose the item answer options for all 16 items. The correct answer options are indicated with an asterisk.

| Item answer option | Proportion of students who chose this answer option in % | | |
|---|---|---|---|
| | Pretest | Posttest | Follow-up test |
| Q1_A | 34.67 | 35.85 | 40.28 |
| Q1_B | 22.92 | 15.41 | 8.33 |
| Q1_C* | 76.50 | 83.02 | 81.94 |
| Q1_D | 17.48 | 12.58 | 9.72 |
| | | | |
| Q2_A | 58.45 | 56.60 | 59.72 |
| Q2_B | 1.72 | 2.20 | 2.78 |
| Q2_C* | 47.85 | 45.28 | 43.06 |
| | | | |
| Q3_A | 8.88 | 7.23 | 5.56 |
| Q3_B | 17.48 | 16.67 | 13.89 |
| Q3_C* | 59.31 | 62.58 | 59.72 |
| Q3_D* | 85.10 | 89.31 | 90.28 |
| | | | |
| Q4_A | 32.95 | 5.97 | 9.72 |
| Q4_B | 26.36 | 25.16 | 18.06 |
| Q4_C | 15.47 | 2.52 | 0.00 |
| Q4_D* | 36.96 | 75.16 | 66.67 |
| Q4_E | 9.46 | 1.57 | 0.00 |
| Q4_F | 16.05 | 15.09 | 19.44 |
| | | | |
| Q5_A | 22.64 | 4.09 | 5.56 |
| Q5_B* | 37.82 | 63.52 | 68.06 |
| Q5_C | 21.20 | 9.12 | 6.94 |
| Q5_D | 24.64 | 25.79 | 22.22 |
| | | | |
| Q6_A | 18.05 | 5.97 | 2.78 |
| Q6_B | 17.48 | 4.72 | 8.33 |
| Q6_C | 14.90 | 16.35 | 18.06 |
| Q6_D* | 49.57 | 73.27 | 70.83 |
| | | | |
| Q7_A | 33.52 | 8.81 | 12.50 |
| Q7_B | 8.60 | 3.14 | 0.00 |
| Q7_C | 19.48 | 5.97 | 9.72 |
| Q7_D* | 48.14 | 91.19 | 91.67 |
| | | | |
| Q8_A | 17.48 | 13.84 | 11.11 |
| Q8_B | 15.47 | 23.90 | 12.50 |
| Q8_C* | 58.17 | 58.18 | 66.67 |
| Q8_D | 16.33 | 8.81 | 11.11 |
| | | | |
| Q9_A | 22.35 | 11.01 | 6.94 |
| Q9_B* | 40.11 | 71.07 | 45.83 |
| Q9_C | 8.60 | 2.83 | 11.11 |
| Q9_D | 9.17 | 2.83 | 1.39 |
| Q9_E | 17.19 | 4.40 | 6.94 |
| Q9_F | 27.22 | 20.44 | 36.11 |
| | | | |
| Q10_A* | 36.68 | 62.26 | 54.17 |
| Q10_B | 43.55 | 24.21 | 26.39 |

CONVERGING LENSES CONCEPT INVENTORY

| Item answer option | Proportion of students who chose this answer option in % | | |
|---|---|---|---|
| | Pretest | Posttest | Follow-up test |
| Q10_C | 24.36 | 14.15 | 20.83 |
| | | | |
| Q11_A* | 36.10 | 62.26 | 50.00 |
| Q11_B | 25.50 | 14.47 | 29.17 |
| Q11_C | 41.83 | 24.53 | 22.22 |
| | | | |
| Q12_A | 5.44 | 3.77 | 1.39 |
| Q12_B* | 25.21 | 49.37 | 50.00 |
| Q12_C | 32.38 | 15.72 | 12.50 |
| Q12_D | 21.49 | 26.73 | 27.78 |
| Q12_E | 38.11 | 15.09 | 19.44 |
| | | | |
| Q13_A | 31.81 | 24.53 | 25.00 |
| Q13_B | 11.75 | 2.83 | 2.78 |
| Q13_C | 37.54 | 43.40 | 37.50 |
| Q13_D* | 24.93 | 35.22 | 41.67 |
| | | | |
| Q14_A | 31.81 | 10.69 | 12.50 |
| Q14_B | 12.03 | 3.77 | 4.17 |
| Q14_C* | 34.67 | 79.87 | 65.28 |
| Q14_D | 35.82 | 11.95 | 22.22 |
| | | | |
| Q15_A | 25.21 | 26.73 | 13.89 |
| Q15_B* | 9.46 | 12.89 | 18.06 |
| Q15_C* | 26.07 | 22.96 | 25.00 |
| Q15_D | 38.40 | 37.11 | 38.89 |
| Q15_E | 19.20 | 17.92 | 13.89 |
| Q15_F* | 26.65 | 36.79 | 33.33 |
| | | | |
| Q16_A | 57.02 | 45.60 | 23.61 |
| Q16_B | 12.89 | 8.81 | 8.33 |
| Q16_C | 23.21 | 9.75 | 5.56 |
| Q16_D* | 12.61 | 39.94 | 62.50 |

**Analyses according to classical test theory:**

TABLE C.II. Values corresponding to the item statistics and the total score reliability.

| Item | Difficulty | Discrimination (16 items) | Discrimination (15 items, w/o Q2) | Alpha-with-item-deleted (16 items) | Alpha-with-item-deleted (15 items, w/o Q2) |
|---|---|---|---|---|---|
| Q1 | .623 | .179 | .181 | .689 | .718 |
| Q2 | .431 | −.032 | n/a | .719 | n/a |
| Q3 | .708 | .235 | .234 | .684 | .713 |
| Q4 | .668 | .393 | .409 | .666 | .695 |
| Q5 | .624 | .448 | .444 | .657 | .689 |
| Q6 | .730 | .340 | .337 | .672 | .703 |
| Q7 | .868 | .291 | .277 | .680 | .710 |
| Q8 | .571 | .272 | .286 | .681 | .709 |
| Q9 | .665 | .190 | .205 | .690 | .718 |
| Q10 | .616 | .400 | .409 | .664 | .694 |
| Q11 | .615 | .377 | .395 | .667 | .695 |
| Q12 | .454 | .372 | .380 | .667 | .697 |
| Q13 | .329 | .205 | .208 | .689 | .718 |
| Q14 | .777 | .402 | .401 | .666 | .696 |
| Q15 | .283 | .260 | .253 | .683 | .712 |
| Q16 | .390 | .361 | .379 | .669 | .697 |

**Analyses according to item response theory:**

Probability functions for all 16 items, P1 (blue) indicates 0 points, P2 (pink) 1 point, and P3 (green) 2 points scored. The axes in each diagram represent the learner's ability (x-axis, theta) and the probability of solving this item correctly (y-axis).

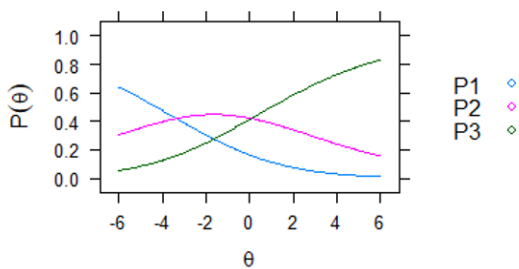

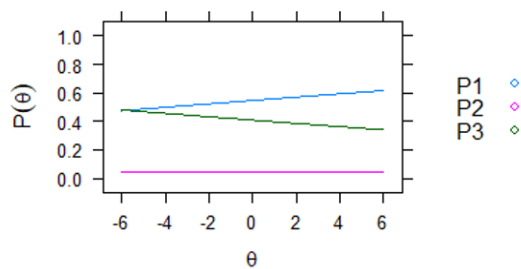

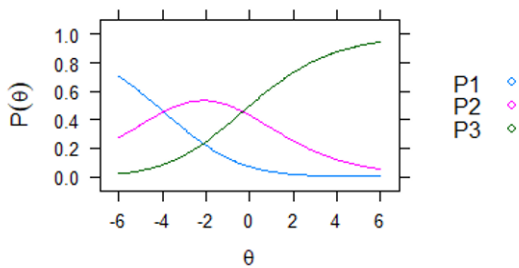

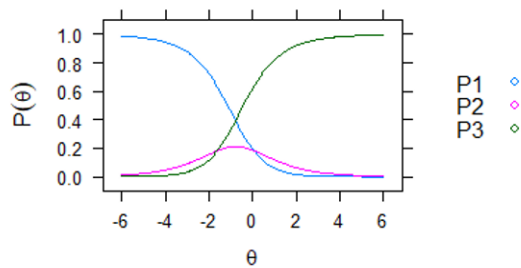

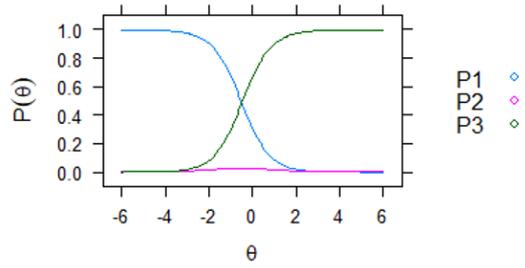
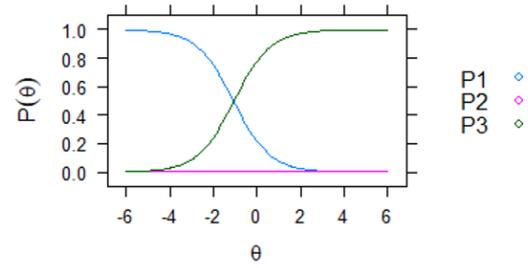
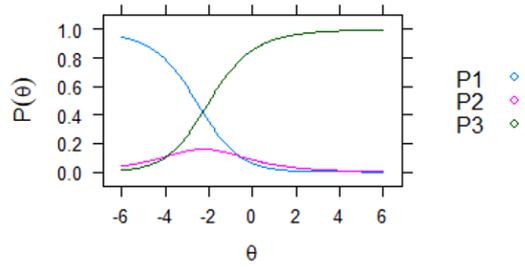
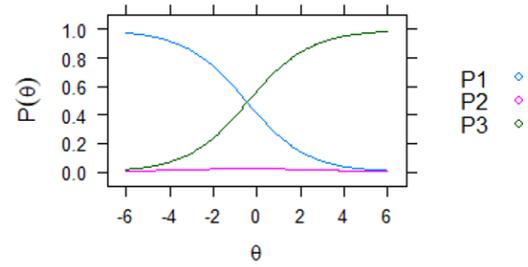
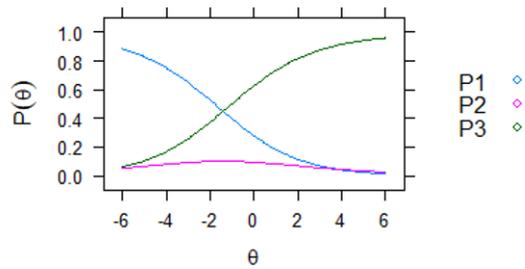
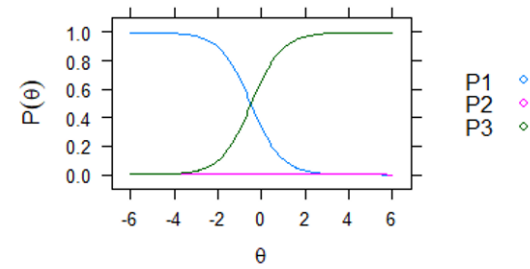
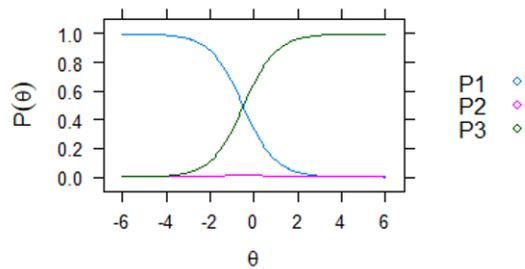
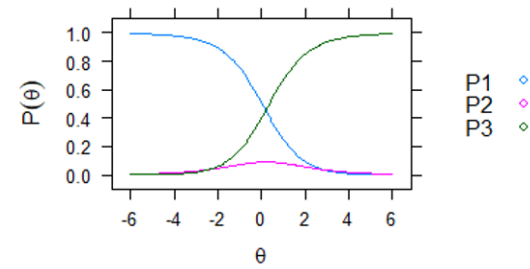

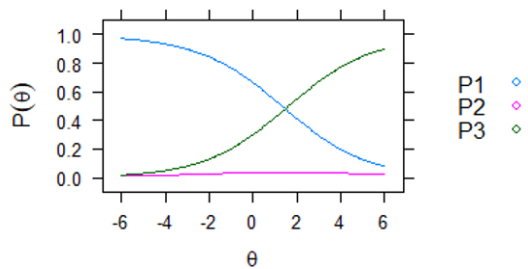
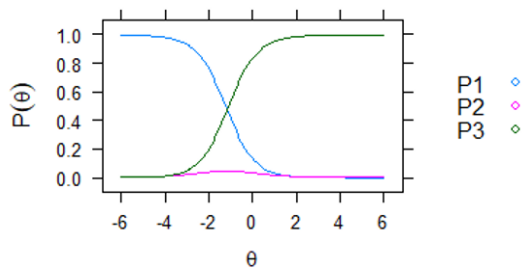
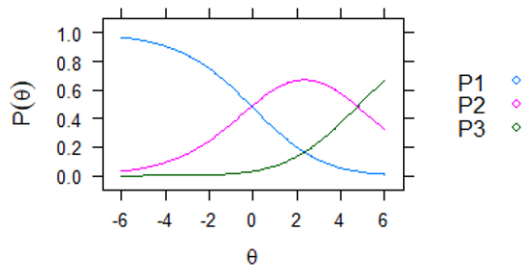
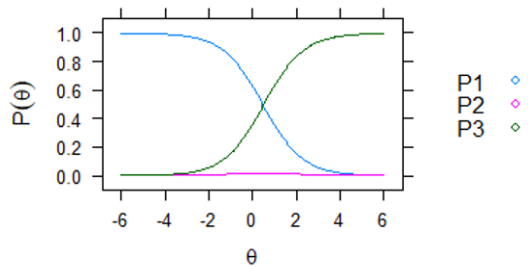

**Structural analyses (Exploratory Factor Analysis):**

Scree Plot obtained from the parallel analysis (factor method: minimum residual solution, eigen values: principal axis factor analysis):

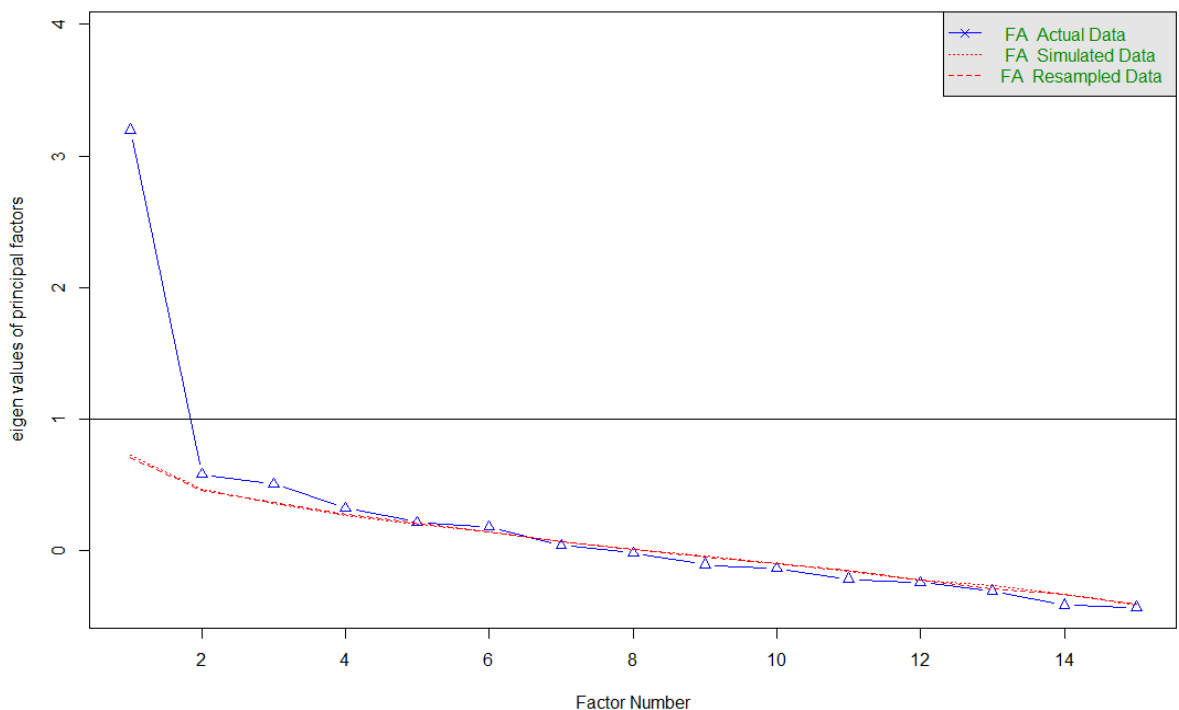

TABLE C.III. Model fits of the tested models in the exploratory factor analysis. A good model should have an RMSR value of close to 0, the RMSEA value should be smaller than 0.05, and the Tucker-Lewis-Index should be above 0.9. Moreover, the structure should make sense "compared to the developers' concept groupings" [9] (p. 488).

| Number of factors | RMSR | RMSEA | Tucker-Lewis-Index | Items with factor loadings < 0.3 | Factors and the items that were associated with them |
|---|---|---|---|---|---|
| 2 | 0.07 | 0.068 | 0.807 | Q9, Q15, Q1, Q7 | MR1: Q14, Q4, Q16, Q12, Q11, Q3, Q10, Q8, Q13<br>MR2: Q6, Q5 |
| 3 | 0.05 | 0.043 | 0.923 | Q13, Q7, Q1, Q15, Q8 | MR1: Q16, Q14, Q4, Q12, Q3, Q9<br>MR2: Q6, Q5<br>MR3: Q11, Q10 |
| 4 | 0.04 | 0.034 | 0.951 | Q8 | MR1: Q12, Q14, Q13, Q3, Q16, Q4, Q15<br>MR2: Q6, Q5<br>MR3: Q10, Q11<br>MR4: Q9, Q7, Q1 |
| 5 | 0.03 | 0.022 | 0.978 | - | MR1: Q14<br>MR2: Q5, Q6<br>MR3: Q10, Q11<br>MR4: Q9, Q4, Q1, Q16, Q7<br>MR5: Q8, Q3, Q15, Q12, Q13 |
| 6 | 0.02 | 0.000 | 1.089 | Q9 | MR1: Q14, Q16, Q12<br>MR2: Q6, Q5<br>MR3: Q11, Q10<br>MR4: Q4<br>MR5: Q1, Q7<br>MR6: Q8, Q15, Q3, Q13 |
| 7 | 0.01 | 0.000 | 1.140 | Q9 | MR1: Q4, Q3<br>MR2: Q1<br>MR3: Q5, Q6<br>MR4: Q11, Q10<br>MR5: Q14, Q12<br>MR6: Q8, Q13, Q15<br>MR7: Q7, Q16 |
| 8 | 0.01 | 0.000 | 1.149 | Q16 | MR1: Q14, Q12<br>MR2: Q1<br>MR3: Q5, Q6<br>MR4: Q11, Q10<br>MR5: Q4, Q9<br>MR6: Q13, Q8, Q15<br>MR7: Q7<br>MR8: Q3 |
| 9 | 0.01 | 0.000 | 1.119 | Q7, Q15 | MR1: Q16, Q4, Q8<br>MR2: Q3<br>MR3: Q9<br>MR4: Q13<br>MR5: Q6, Q5<br>MR6: Q10, Q11<br>MR7: Q1<br>MR8: Q12<br>MR9: Q14 |